\documentclass[
 aip,
 jcp,
amsfonts,
amsmath,
amssymb,
reprint,%
]{revtex4-1}

\usepackage{bm}
\usepackage{graphicx}
\usepackage[utf8]{inputenc}
\usepackage[T1]{fontenc}      
\usepackage{array}
\usepackage{tablefootnote}
\usepackage{multirow}
\usepackage{subfigure}
\usepackage[colorlinks,linkcolor=blue,urlcolor=blue,citecolor=blue]{hyperref}

\DeclareMathOperator{\Tr}{Tr}
\DeclareMathOperator{\kzero}{\mathbf{K_0}}
\DeclareMathOperator{\br}{\mathbf{\widetilde{R}}}
\DeclareMathOperator{\Hm}{\mathbf{H}}
\DeclareMathOperator{\M}{\mathbf{M}}
\DeclareMathOperator{\bv}{\mathbf{\widetilde{V}_{H}}}
\DeclareMathOperator{\bvxc}{\mathbf{V_{xc}}}

\DeclareMathOperator{\F}{\mathbf{F}}

\DeclareMathOperator{\bx}{\mathbf{x}}
\DeclareMathOperator{\bxc}{\mathbf{xc}}
\DeclareMathOperator{\blambda}{\mathbf{\Lambda}}

\DeclareMathOperator{\LL}{\mathbf{L}}

\begin{document}

\title{Spectral finite-element formulation of the optimized effective potential method for atomic structure in the random phase approximation}
\author{Shubhang Krishnakant Trivedi}
\affiliation{College of Engineering, Georgia Institute of Technology, Atlanta, GA 30332, USA}
\author{Phanish Suryanarayana}
\email[Email: ]{phanish.suryanarayana@ce.gatech.edu}
\affiliation{College of Engineering, Georgia Institute of Technology, Atlanta, GA 30332, USA}
\affiliation{College of Computing, Georgia Institute of Technology, Atlanta, GA 30332, USA}

\begin{abstract}
We present a spectral finite-element formulation of the optimized effective potential (OEP) method for atomic structure calculations in the random phase approximation (RPA). In particular, we develop a finite-element framework that employs a polynomial mesh with element nodes placed according to the Chebyshev--Gauss--Lobatto scheme, high-order $\mathcal{C}^0$-continuous Lagrange polynomial basis functions, and Gauss--Legendre quadrature for spatial integration. We employ distinct polynomial degrees for the orbitals, Hartree potential, and RPA--OEP exchange--correlation potential. Through representative examples, we verify the accuracy of the developed framework, assess the fidelity of one-parameter double-hybrid functionals constructed with RPA correlation, and develop a machine-learned model for the RPA--OEP exchange--correlation potential at the level of the generalized gradient approximation, based on the kernel method and linear regression.
\end{abstract}

\maketitle

\section{Introduction}\label{Introduction}

Over the past several decades, Kohn--Sham density functional theory (DFT) \cite{Kohn-shampaper, Hohenberg-Kohnpaper}  has become a cornerstone  in materials and chemical sciences research, owing to the fundamental physical insight it can provide and its strong predictive power. Grounded in the first-principles of quantum mechanics, Kohn-Sham DFT has been widely adopted since it offers an effective balance between conceptual simplicity, broad applicability, and a favorable accuracy-to-computational-cost ratio relative to other ab initio methods. Nevertheless, although substantially less expensive than wavefunction-based approaches, Kohn--Sham DFT remains computationally demanding, which continues to limit the size and complexity of systems that can be investigated.

The accuracy and computational cost of Kohn--Sham DFT calculations  are primarily dictated by the choice of exchange--correlation, which encodes many-body electron interactions and constitutes the principal approximation within the Kohn--Sham formalism. Owing to the absence of a universal exchange--correlation functional, a hierarchy of approximations has been established and organized within Jacob’s ladder \cite{jacobladderperdew}, wherein successive rungs generally offer improved accuracy at the expense of increased computational cost. The lowest four rungs, encompassing local, semilocal, and hybrid functionals, are the most widely used in practice. The fifth and highest rung consists of nonlocal many-body correlation methods based on the adiabatic-connection fluctuation--dissipation (ACFD) theorem \cite{ACFDperdew}, in which exchange is treated exactly and the correlation energy is computed from the density response function. This requires access to unoccupied orbitals and yields improved accuracy at a substantially higher computational cost.

The fifth rung of Jacob’s ladder includes the random phase approximation (RPA) exchange--correlation, which  is capable of capturing van der Waals interactions, eliminating self-interaction errors, and accurately describing both small-gap and metallic systems, thereby enabling benchmark-level accuracy for condensed-matter systems \cite{ren2012random, eshuis2012electron}.  Even in non-self-consistent form, RPA has been shown to provide improved predictive performance over lower-rung functionals across a broad range of properties, including surface energies, adsorption energies, binding energies, cohesive energies, and lattice constants \cite{RenRPA3, KresseRPAlattice, Hutterliquidwater, JiangRPAstability, Tkatchenkovanderwaals, ThygesenAdsorptionenergies, DierRPA, RPA-SELF-CONSISTENT-PBE0-PITTS-PSEUDOPOTENTIAL-SOLIDS}. However, the associated computational cost is orders of magnitude higher than that of local/semilocal exchange--correlation \cite{KresseRPAForces, boqinRPA, ShikharRPA}, which has limited its practical use, particularly in self-consistent form that necessitates the implementation of the optimized effective potential (OEP) method \cite{oepjdtalman,Sham-Schluter-equation-OEP}. More broadly, the substantial cost and algorithmic complexity have hindered the systematic testing and development of fifth-rung exchange--correlation, as well as the generation of high-quality training data for machine-learned models.

Atomic structure calculations \cite{bhowmik2025spectral, dftatom, CERTIK2024109051, lehtolafem, lehtolafem2, lehtolafem3, lehtolafem4, Romanowski1, Romanowski2, OZAKI20111245, numerical_grid_atom, herman1963atomic, oepjdtalman, LDAexpatom, secondorderKSMP2, gwasphericalatomshellgren, vacondiopaper, Uzulis_2022} exploit the spherical symmetry of isolated atoms to solve the electronic-structure problem in radial coordinates, providing an attractive  setting for assessing the effectiveness and guiding the development of new exchange--correlation. In particular, high-quality reference data from coupled-cluster [CCSD(T)] \cite{CCSD(T)inversion,CCSD(T)openshellatoms}, configuration-interaction (CI) \cite{chakravarthyhelium, chakravarthyberylliumandothers}, and quantum Monte Carlo (QMC) \cite{umrigargonze} calculations are readily available for atoms, enabling rigorous and systematic benchmarking. This availability has, in turn, motivated the implementation of fifth-rung functionals within the OEP formalism, including MP2- and RPA-based variants, employing Gaussian, numerical, or cubic-spline basis representations \cite{secondorderKSMP2, OEP-correlation-2-ivanov, OEP-correlation-4-ivanov-sc2, OEP-correlation-5-grabowski, OEP-correlation-6-ivanov, OEP-correlation-7-ivanov, OEP-correlation-8-Schweigert, gwasphericalatomshellgren, vermabarlettpaper, OEP-correlation-RPA-1-hellgren-molecules, vacondiopaper, scRPAgorlingpaper}. Moreover, the calculations are orders of magnitude more efficient than their three-dimensional counterparts. In contrast to implementations for extended systems, which are typically restricted to pseudopotential calculations owing to their prohibitive computational cost, atomic structure calculations are highly efficient and therefore permit all-electron treatments across the entire periodic table. Although such calculations are restricted to isolated atoms, they nonetheless provide a valuable platform for generating high-fidelity data for machine-learned models \cite{realspacesecondorderML, MLCI}, particularly when strategies to emulate diverse chemical environments relevant to molecules and extended systems are incorporated into the atomic framework. This provides the motivation for the present effort.

In this work, we introduce a spectral finite-element formulation of the OEP method for atomic structure calculations with RPA exchange-correlation. In particular, we develop an adaptive, spectral finite-element framework that employs distinct interpolation degrees for different quantities. Using representative examples, we demonstrate the accuracy of the developed framework, assess the fidelity of one-parameter double-hybrid functionals constructed with RPA correlation, and develop a machine-learned model for the RPA--OEP exchange--correlation potential at the GGA level, based on the  kernel method and linear regression.

The remainder of this manuscript is organized as follows. In Section~\ref{Sec:AtomicStructure}, we describe the RPA--OEP atomic structure formalism.  In Section~\ref{Sec:Implementation}, we describe the spectral finite-element framework. In Section~\ref{Sec:Results}, we verify its accuracy and apply it to develop exchange-correlation functionals. Finally, we provide concluding remarks in Section~\ref{Sec:Conclusions}.


\section{RPA--OEP atomic structure formalism} \label{Sec:AtomicStructure}
Consider a charge neutral and closed-shell isolated atom with atomic number $Z$. The generalized Kohn--Sham DFT energy functional in radial coordinates takes the form: 
\begin{align}
    E[\widetilde{R}_{nl},\lambda_{nl}] = T_s[\widetilde{R}_{nl}] + E_{xc}[\widetilde{R}_{nl},\lambda_{nl}]+ E_{el}[\rho]\ , \label{Eq:energy}
\end{align}
where $T_s$ is the kinetic energy of non-interacting electrons, $E_{xc}$ is the exchange--correlation energy, and $E_{el}$ is the electrostatic energy; $n$, $l$, and $m$ are the principal, azimuthal, and magnetic quantum numbers, respectively; $\widetilde{R}_{nl}/r $ is the radial component of the Kohn--Sham orbital, the corresponding eigenvalue and occupation being $\lambda_{nl}$ and $C_l f_{nl}$, respectively;  and $\rho$ is the electron density:
\begin{equation}
    \rho(r) = \frac{1}{2\pi r^2}\sum_{nl} C_{l}f_{nl} \widetilde{R}_{nl}^2(r) \,,  \label{Eq:density}
\end{equation}
with $C_l = 2 l+1$ and $f_{nl} \in \{0,1\}$.  Considering the RPA exchange--correlation, the energy functional components take the form \cite{CERTIK2024109051,  bhowmik2025spectral, cinal, RPAcorrelationenergyjiangengel}:
\begin{widetext}
\begin{subequations}
\begin{align}
 T_s[\widetilde{R}_{nl}]  = & -\sum_{nl} C_{l}f_{nl} \int \left(\widetilde{R}_{nl}(r)\frac{d^2\widetilde{R}_{nl}(r)}{dr^2}  - \frac{l(l+1)}{r^2} \widetilde{R}_{nl}^2(r) \right) \, dr \label{Eq:KE} \ , \\
 E_{el}[\rho]  = &  \max_{\widetilde{V}_{H}} \left[\int \left( -\frac{1}{2} \left(\frac{d \widetilde{V}_{H}(r)}{dr}\right)^2 + 4\pi r \widetilde{V}_{H}(r) \rho(r) \right)dr\right] - 4\pi Z \int r\rho(r)   dr  \,, \label{Eq:ElecE} \\
E_{xc}[\widetilde{R}_{nl},\lambda_{nl}]   =&  - \sum_{nl,n^\prime l^\prime }C_l f_{nl} C_{l^\prime}f_{n^\prime l^\prime} \sum_{l'\!'} \begin{pmatrix}
        l & l' & l'\!' \\
        0 & 0  & 0
      \end{pmatrix}^{2} \nonumber
\iint \widetilde{R}_{nl}(r)\widetilde{R}_{n'l'}(r)\nu_{l'\!'} (r,r^\prime) \widetilde{R}_{nl}(r')\widetilde{R}_{n'l'}(r') \, dr dr^\prime \nonumber \\
&  + \frac{1}{2\pi}\sum_{l'\!'}C_{l'\!'}\int \Tr\Big[K_{l'\!'}(i\omega) \nu_{l'\!'}+ \log(I - K_{l'\!'}(i\omega)\nu_{l'\!'})\Big] \, d\omega \,, \label{Eq:Exc}
\end{align} 
\end{subequations}
\end{widetext}
where $\widetilde{V}_H/r$ is the Hartree potential, $\left(\begin{smallmatrix} l&l^\prime&l'\!'\\ \\ 0&0&0 \end{smallmatrix}\right)$ is the Wigner-3j symbol, $\nu_{l'\!'}$ is the radial Coulomb operator, and $K_{l'\!'}(i\omega)$ is the radial density response function at imaginary frequency $i \omega$, $I$ is the identity operator, and the limits of integration  here and throughout, if not specified, are  $0$ to $\infty$. The exchange--correlation energy has been decomposed into the exact exchange ($E_X$) and RPA correlation energy ($E_c$) contributions. In defining the correlation energy, we have used the notation: $AB = \int A(r,r^\prime)B(r^\prime,r'\!') \, dr^\prime$  and $\Tr[AB] = \iint A(r,r^\prime)B(r,r^\prime) \, dr dr^\prime$, for given integral operators $A$ and $B$.   The radial Coulomb operator takes the form:
\begin{align}
\nu_{l'\!'} (r,r^\prime )  = \frac{[\min(r,r^\prime)]^{l'\!'}}{[\max(r,r^\prime)]^{l'\!'+1}} \,, \label{Eq:nu}
\end{align}
and radial density response function can be written as:
\begin{widetext}   
\begin{align}
    K_{l'\!'}(r,r';i\omega)  = 2\sum_{nl,n'l'}\frac{(f_{nl}-f_{n'l'})}{\lambda_{nl}-\lambda_{n'l'}+i\omega}\frac{C_lC_{l'}}{C_{l'\!'}}\begin{pmatrix}
        l & l' & l'\!' \\
        0 & 0  & 0
      \end{pmatrix}^{2}\widetilde{R}_{nl}(r)\widetilde{R}_{n'l'}(r)\widetilde{R}_{nl}(r')\widetilde{R}_{n'l'}(r') \,. \label{eq:Chi0}
\end{align}   
\end{widetext}

The variation of the exchange--correlation energy with respect to $\widetilde{R}_{nl}$ results in a nonlocal potential that depends on the orbitals as well as the eigenvalues, which makes the solution for the electronic ground state particularly challenging. To overcome this, the OEP formalism approximates the nonlocal exchange--correlation potential operator with a local/multiplicative potential \cite{OEP-exchange-4-kummel-atoms-molecules, OEP-correlation-GL2-2, oepjdtalman, engel_dreizler_2011}. In particular, the  local potential is written as the solution to the following variational problem \cite{oepjdtalman}:
\begin{align}
 \min_{V_s} \; &E[\widetilde{R}_{nl},  \lambda_{nl}] \nonumber \\ 
 &\mathrm{s.t.} \nonumber\\  
 \Bigg[-\frac{1}{2}\frac{d^2}{dr^2}+\frac{l(l+1)}{2r^2} + &V_s(r)\Bigg]\widetilde{R}_{nl}(r) = \lambda_{nl}\widetilde{R}_{nl}(r) \,,
 \label{eq:EOEPmin}   
\end{align}
where $V_s$ can be  decomposed as follows:
\begin{align}
V_s(r) =-\frac{Z}{r} + \frac{\widetilde{V}_H(r)}{r} + V_{xc}(r) \,, \label{eq:Vs}
\end{align}
with $V_{xc}$ being the OEP exchange--correlation potential.  The energy functional $E$ should therefore be stationary with respect to the variation in potential $V_s$:
\begin{align}
\frac{\delta {E}}{\delta V_{s}(r)}\! &= \!
\sum_{nl} \left[\int\! \frac{\delta {E}}{\delta \widetilde{R}_{nl}(r')} \frac{\delta \widetilde{R}_{nl}(r^\prime)}{\delta V_{s}(r)} dr^\prime \! +\! \frac{\delta {E}}{\delta \lambda_{nl}} \frac{\delta \lambda_{nl}}{\delta V_{s}(r)}\right] \!=\! 0 \,, \label{eq:dEdVs} 
\end{align}  
while simultaneously satisfying the constraint in Eq.~\ref{eq:EOEPmin}. Using this constraint, it can be shown that:
\begin{align} 
 \frac{\delta {E}}{\delta \widetilde{R}_{nl}(r^\prime)} &=  \Bigg(4C_l  f_{nl}\Big(\lambda_{nl} - V_{xc}(r^\prime)\Big) + \widehat{V}_{nl}\Bigg)\widetilde{R}_{nl}(r^\prime) \, , \label{eq:dEdR}
\end{align} 
where the nonlocal potential associated with the exchange--correlation energy can be derived to be: 
\begin{widetext}
\begin{align}
    \widehat{V}_{nl} \widetilde{R}_{nl}(r') = & - 4 C_lf_{nl} \sum_{n^\prime l^\prime} C_{l^\prime} f_{n^\prime l^\prime}\sum_{l'\!'} \begin{pmatrix}
        l & l' & l'\!' \\
        0 & 0  & 0
      \end{pmatrix}^{2} \nonumber \widetilde{R}_{n'l'}(r^\prime) \int \widetilde{R}_{n'l'}(r^\prime \!^\prime)\widetilde{R}_{nl}(r^\prime \!^\prime)\nu_{l'\!'} (r^\prime \!^\prime,r^\prime) \, dr^\prime \!^\prime   \\ 
       -\frac{2 C_l}{\pi} & \sum_{n^\prime l^\prime}C_{l^\prime}(f_{nl}-f_{n'l'})\sum_{l'\!'} \begin{pmatrix}
        l & l' & l'\!' \\
        0 & 0  & 0
      \end{pmatrix}^{2}\widetilde{R}_{n'l'}(r')\int_{-\infty}^{\infty} \Bigg(\int \frac{\widetilde{R}_{n'l'}(r^\prime \!^\prime)\widetilde{R}_{nl}(r^\prime \!^\prime)}{ \lambda_{nl}-\lambda_{n'l'}+i\omega }  W_{l'\!'}^{c}(r^\prime \!^\prime,r^\prime;i\omega) \, dr^\prime \!^\prime\Bigg) \, d\omega \,,  \label{eq:Vxc}
\end{align} 
\end{widetext} 
the first term representing the exact exchange operator ($\widehat{V}_{X}$) and the second term representing the RPA correlation operator ($\widehat{V}_{c}$), with the correlation part of the screened Coulomb interaction given by the relation:
\begin{align}
    W^c_{l'\!'}  =&  \nu_{l'\!'}\left( (I - K_{l'\!'}(i\omega)\nu_{l'\!'})^{-1}   -  I\right) \,. \label{eq:Wc}
\end{align}
It follows from perturbation theory that:
\begin{subequations} \label{Eq:Variations}
\begin{align} 
 \frac{\delta \widetilde{R}_{nl}(r^\prime)}{\delta V_{s}(r)}  & = -G_{nl}(r,r^\prime)\widetilde{R}_{nl}(r) \,, \label{eq:dRdVs} \\
\frac{\delta \lambda_{nl}}{\delta V_s(r)} & = \widetilde{R}_{nl}^2 (r) \,,\label{eq:deigdVs}
\end{align}
\end{subequations}
where the radial static Green's function is given by:
\begin{align}
    G_{nl}(r,r^\prime) = \sum_{n^\prime  \neq n}  \frac{\widetilde{R}_{n'l}(r)\widetilde{R}_{n'l}(r')}{\lambda_{n'l}-\lambda_{nl}} \,, \label{eq:Greenfunction}
\end{align}
Inserting Eqs.~\ref{eq:dEdR}, \ref{Eq:Variations}, and \ref{eq:Greenfunction} into Eq.~\ref{eq:dEdVs} and rearranging leads to the equation for the OEP exchange–-correlation potential:
\begin{align}
\int K_{0}(r,r^\prime;0)V_{xc}(r^\prime) \, dr^\prime = \Lambda_{xc}(r) \,,\label{eq:OEP} 
\end{align}
where the static response function $K_{0}(r,r^\prime;0)$ is as defined in Eq.~\ref{eq:Chi0}:
\begin{align}
    K_0(r,r';0) = -4\sum_{nl}f_{nl} C_l \widetilde{R}_{nl}(r) G_{nl}(r,r')\widetilde{R}_{nl}(r') \,, \label{eq:Chi0static} 
\end{align}
and the right hand side function:
\begin{align}
\Lambda_{xc}(r) = \sum_{nl} \Bigg[  - \widetilde{R}_{nl}(r) \int  &G_{nl}(r,r')   \widehat{V}_{nl} \widetilde{R}_{nl}(r^\prime)\, dr^\prime \nonumber \\
 & + \frac{\delta E_{xc}}{\delta \lambda_{nl}}\widetilde{R}_{nl}^2(r)\Bigg], \label{eq:OEPRHS}
\end{align}    
with the radial static Green's function $G_{nl}(r,r')$ given by Eq.~\ref{eq:Greenfunction},  $\widehat{V}_{nl} \widetilde{R}_{nl}(r')$ given by Eq.~\ref{eq:Vxc}, and 
\begin{widetext}
    \begin{align}
    \frac{\delta E_{xc}}{\delta \lambda_{nl}} =& \frac{1}{\pi} \sum_{n' l'}C_{l}C_{l'}(f_{nl}-f_{n'l'})\sum_{l'\!'} \begin{pmatrix}
        l & l' & l'\!' \\
        0 & 0  & 0
      \end{pmatrix}^{2}\int_{-\infty}^{\infty} \Bigg(\iint \frac{\widetilde{R}_{nl}(r^\prime)\widetilde{R}_{n'l'}(r^\prime)\widetilde{R}_{nl}(r^\prime \!^\prime)\widetilde{R}_{n'l'}(r^\prime \!^\prime) }{(\lambda_{nl}-\lambda_{n'l'}+i\omega)^2} W_{l'\!'}^{c}(r^\prime \!^\prime,r^\prime;i\omega)\, dr^\prime dr^\prime \!^\prime \Bigg)\, d\omega  \,. \label{eq:dExcdeig}
\end{align}
\end{widetext}
The OEP exchange--correlation potential can be determined from Eq.~\ref{eq:OEP} up to an additive constant, owing to the singular nature of \(K_0(r,r';0)\). The electronic ground state for the atomic structure problem in RPA--OEP can therefore be determined by the self-consistent solution of the following angular momentum dependent eigenproblems:
\begin{widetext}
\begin{align}
\Bigg[\mathcal{H}_l \equiv -\frac{1}{2}\frac{d^2}{dr^2} + \frac{l(l+1)}{2r^2}  - \frac{Z}{r} + \frac{\widetilde{V}_H(r)}{r} + V_{xc}(r) \Bigg]\widetilde{R}_{nl}(r) = \lambda_{nl}\widetilde{R}_{nl}(r) \,, \quad \widetilde{R}_{nl}(0) = 0 \,, \widetilde{R}_{nl}(\infty) = 0 \,, \label{Eq:Eigenproblem}
\end{align}
\end{widetext}
where $\widetilde{V}_H$ is the solution to the Poisson equation:
\begin{align}
- \frac{d^2 \widetilde{V}_H(r)}{d r^2} = 4 \pi r \rho(r),    \quad
         \widetilde{V}_{H}(0)= 0\,, \widetilde{V}_{H}(\infty) = Z \,,  \label{eq:Poisson}
\end{align}
the electron density $\rho$ is given by Eq.~\ref{Eq:density}, and the OEP exchange--correlation potential $V_{xc} = V_x + V_c$ is the solution to Eq.~\ref{eq:OEP}.

The formalism presented above assumes the electron density to be spherically symmetric, as is standard in atomic structure calculations. While this symmetry is preserved in the absence of external fields and is recovered in many-body wavefunction theories, the mean-field nature of Kohn--Sham DFT implies that open-shell atoms do not, in general, exhibit spherical symmetry. This limitation can be remedied by allowing fractional occupations, \( f_{nl} \in [0,1] \), chosen to be identical for all states within a given open shell while preserving the total number of electrons in that shell. With this modification, the above formalism is  applicable to open-shell atoms.
\section{Spectral Finite-element Framework} \label{Sec:Implementation}
\begin{figure*}[htbp]
\centering
\includegraphics[width=0.9\textwidth]{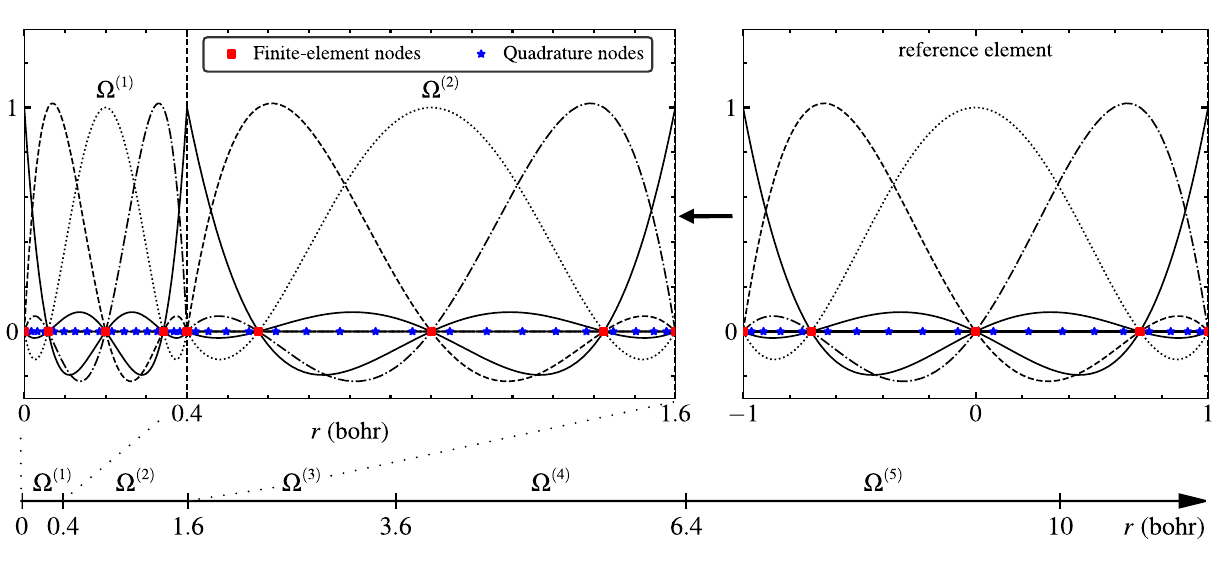}
\caption{\label{Fig:FEM} Illustration of the spectral finite-element framework developed for RPA--OEP atomic-structure calculations. The example shown uses a domain size of $R_{max} = 10$ bohr, $N_e=5$ elements, polynomial degree $p_1=p_2=p_3=4$ for interpolation, and a quadrature order of $20$.}
\end{figure*}

We develop a spectral finite-element framework for the RPA$-$OEP atomic structure formalism  presented in the previous section. The choice of the finite element method is motivated by its systematic improvability, ability to employ high-order approximations, and flexibility in accommodating adaptive grids, which together makes the scheme accurate as well as efficient in the present all-electron OEP context. The spectral finite-element framework developed in this work, schematically illustrated in Fig.~\ref{Fig:FEM}, is described in detail below. Indeed, the term \emph{spectral} refers to the use of high-order polynomial approximations in conjunction with appropriately chosen quadrature rules, consistent with common usage in the finite-element literature \cite{PATERA1984468}.

The radial domain $\Omega$ is restricted to $r \in [0, R_{\max}]$, justified by the exponential decay of the orbitals. This domain is then partitioned into subdomains \( \{ \Omega^{(e)} \}_{e=1}^{N_e} \), referred to as elements, where \( N_e \) is the total number of elements. In each element, we adopt the following basis set expansions for $r \in \Omega^{(e)}$:
\begin{subequations}
\begin{align}
\widetilde{R}_{nl}(r) & = \sum_{j=0}^{p_1}  [\br_{nl}]^{(e)}_{j} \phi^{p_1(e)}_{j}(r) \,, \label{eq:Rtildediscretization} \\
\widetilde{V}_{H}(r) & = \sum_{j=0}^{p_2}  [\bv]^{(e)}_{j} \phi_{j}^{p_2(e)}(r) \,, \label{eq:Vhtildediscretization} \\
V_{xc}(r) & = \sum_{j=0}^{p_3}  [\bvxc]^{(e)}_{j} \phi_{j}^{p_3(e)}(r) \,, \label{eq:Vxcdiscretization}
\end{align}
\end{subequations}
where \( [\,\cdot\,]_{j} \) denotes the value of the quantity at the \( j \)-th node. In addition, \( p_1 \), \( p_2 \), and \( p_3 \) denote the degrees of the polynomial basis functions \( \phi^{p_1}_{j} \), \( \phi^{p_2}_{j} \), and \( \phi_{j}^{p_3} \), respectively, each of which is compactly supported on \( \Omega^{(e)} \) and satisfies the Kronecker delta property. Indeed, the number of nodes in each element is one greater than the polynomial degree, i.e., there are $p_1+1$, $p_2+1$, and $p_3+1$ nodes in the elements for $\widetilde{R}_{nl}$, $\widetilde{V}_{H}$, and $V_{xc}$, respectively. The motivation of a different (higher) degree for $\widetilde{V}_{H}$ relative to $\widetilde{R}_{nl}$ is to increase the efficiency of the calculations, the former having higher frequency content than the latter. Indeed, the solution of the eigenproblem associated with $\widetilde{R}_{nl}$ is significantly more  expensive than the solution of the linear system associated with $\widetilde{V}_{H}$. The motivation of a different (lower) degree for $V_{xc}$ relative to both $\widetilde{R}_{nl}$ and $\widetilde{V}_{H}$ is for the purposes of numerical stability. Here and henceforth, the lowercase indices represent the local numbering of the nodes within each element, while the uppercase indices will denote the global numbering. The mapping between them in the  elements used for $\widetilde{R}_{nl}$, $\widetilde{V}_{H}$, and $V_{xc}$, is denoted by \( \mathcal{M}^{p_1} \), \( \mathcal{M}^{p_2} \), and \( \mathcal{M}^{p_3} \), respectively, with \( \mathcal{M}^{p_1(e)}_{k} \), \( \mathcal{M}^{p_2(e)}_{k} \), and \( \mathcal{M}^{p_3(e)}_{k} \) providing the global index of the node with local index \( k \) in the \( e \)-th element. In what follows, we present the the discrete (weak) form of the three main equations within the RPA$-$OEP formalism, namely, Eqs.~\ref{Eq:Eigenproblem}, Eq.~\ref{eq:Poisson}, and \ref{eq:OEP}. 

The discrete form of the angular momentum dependent eigenproblem in Eq.~\ref{Eq:Eigenproblem} can be written as:
\begin{align}
\Hm_{l} \! \br_{nl} &= \lambda_{nl} \! \M \! \br_{nl}  \,, \label{eq:GenEigenProblem}
\end{align}
where the global matrices/vectors are assembled from the element counterparts  as: 
\begin{subequations}
\begin{align}
    [\Hm_l]_{IJ} &= \sum_{e=1}^{N_{e}}\sum_{i, j = 0}^{p_1}[\Hm_l]^{(e)}_{ij} \delta_{I\mathcal{M}^{1(e)}_{i}}\delta_{J\mathcal{M}^{1(e)}_{j}} \,, \label{eq:assembled_hamiltonian}\\
    [\M]_{IJ} &= \sum_{e=1}^{N_{e}}\sum_{i, j = 0}^{p_1}[\M]^{(e)}_{ij} \delta_{I\mathcal{M}^{1(e)}_{i}}\delta_{J\mathcal{M}^{1(e)}_{j}} \,, \label{eq:assembled_overlap} \\
     [\br_{nl}]_{I} &= \sum_{e=1}^{N_{e}}\sum_{i = 0}^{p_1} g_i^{(e)} [\br_{nl}]^{(e)}_{i} \delta_{I\mathcal{M}^{1(e)}_{i}}\,, \label{eq:assembled_rtilde} 
\end{align}    
\end{subequations} 
where \(\delta\) denotes the Kronecker delta, \(g_i^{(e)} = 1\) for nodes that are not shared between elements and \(g_i^{(e)} = 0.5\) for nodes shared between adjacent elements, and the element matrices take the form:
\begin{widetext}
\begin{subequations}
\begin{align}
   [\Hm_l]^{(e)}_{ij} & =  \int \Bigg[\frac{1}{2}\frac{d\phi^{p_1(e)}_{i}(r)}{dr} \frac{d \phi^{p_1(e)}_{j}(r)}{d r} + \phi^{p_1(e)}_{i}(r)\Bigg(\!\frac{l(l+1)}{2r^2}\!\Bigg)\phi^{p_1(e)}_{j}(r) +\phi^{p_1(e)}_{i}(r)\Bigg( \!\!-\frac{Z}{r} + \frac{\widetilde{V}_H(r)}{r} +\! V_{xc}(r)\!\Bigg)\phi^{p_1(e)}_{j}(r)\Bigg]\, dr \,,  \label{eq:element_hamiltonian}\\
[\M]^{(e)}_{ij} & =  \int \phi^{p_1(e)}_{i}(r) \phi^{p_1(e)}_{j}(r) \, dr \,.  \label{eq:element_overlap}   
\end{align}    
\end{subequations}
\end{widetext}
The homogeneous Dirichlet boundary conditions on \( \widetilde{R}_{nl} \) are enforced by removing the first and last rows and columns of the matrices \( \Hm_{l} \) and \( \M \), along with the first and last entries of the vector \( \br_{nl} \). 

The discrete form of the Poisson problem in Eq.~\ref{eq:Poisson} can be written as:
\begin{align}
\LL \! \bv = \F \label{eq:FEPossion} 
\end{align}
where the global matrices/vectors are assembled from the element counterparts as: 
\begin{subequations}
\begin{align}
    [\LL]_{IJ} &= \sum_{e=1}^{N_{e}}\sum_{i,j=0}^{p_2}[\LL]^{(e)}_{ij} \delta_{I\mathcal{M}^{p_2(e)}_{i}}\delta_{J\mathcal{M}^{p_2(e)}_{j}} \,, \label{eq:assembled_Laplacian} \\
    [\F]_{I} &=   \sum_{e=1}^{N_{e}}\sum_{i=0}^{p_2}[\F]^{(e)}_{i} \delta_{I\mathcal{M}^{p_2(e)}_{i}} \,,  \label{eq:assembled_PoissonRHS} \\
   [\bv]_{I} &=   \sum_{e=1}^{N_{e}}\sum_{i=0}^{p_2}g_i^{(e)} [\bv]^{(e)}_{i} \delta_{I\mathcal{M}^{p_2(e)}_{i}} \,, \label{eq:assembled_vhtilde}
\end{align}
\end{subequations}
with the element matrices/vectors taking the form:
\begin{subequations}
\begin{align}
    [\LL]^{(e)}_{ij} &= \int \frac{d\phi_{i}^{p_2(e)}(r)}{dr} \frac{d \phi_{j}^{p_2(e)}(r)}{d r} \, dr\, ,\label{eq:element_Laplacian} \\ 
    [\F]^{(e)}_{i} &= 4\pi \int r \rho(r) \phi_{i}^{p_2(e)} (r)\, dr \, .\label{eq:element_PoissonRHS} 
\end{align}    
\end{subequations}
Above, the density $\rho$ is calculated using Eq.~\ref{Eq:density}, while using the basis expansion of $\widetilde{R}_{nl}$ (Eq.~\ref{eq:Rtildediscretization}). The homogeneous Dirichlet boundary condition on \( \widetilde{V}_{H} \) at \( r = 0 \) is enforced by removing the first row and column of the matrix \( \LL \), together with the first entries of the vectors \( \bv \) and \( \F \). The nonhomogeneous Dirichlet boundary condition at \( r = R_{\max} \) is imposed by removing the last row and column of \( \LL \) and the last entries of \( \bv \) and \( \F \), with  \( \F \) updated accordingly to preserve the solution.

The discrete form of the OEP equation in Eq.~\ref{eq:OEP} can be written as:
\begin{align}
   \kzero \! \bvxc = \blambda_{\bxc} \,, \label{eq:FEOEP}
\end{align}
where the global matrices/vectors are assembled from the element counterparts as: 
\begin{subequations}
    \begin{align}
        [\kzero]_{IJ} &= \sum_{e, e'=1}^{N_{e}}\sum_{i, j=0}^{p_3}[\kzero]^{(e, e')}_{ij} \delta_{I\mathcal{M}^{p_3(e)}_{i}}\delta_{J\mathcal{M}^{p_3(e')}_{j}} \,, \label{eq:assembled_chi0}\\
        [\blambda_{\bxc}]_{I} &= \sum_{e=1}^{N_{e}}\sum_{i=0}^{p_3}[\blambda_{\bxc}]^{(e)}_{i} \delta_{I\mathcal{M}^{p_3(e)}_{i}} \,, \label{eq:assembled_OEPRHS} \\
         [\bvxc]_{I} &= \sum_{e=1}^{N_{e}}\sum_{i=0}^{p_3} g_i^{(e)} [\bvxc]^{(e)}_{i} \delta_{I\mathcal{M}^{p_3(e)}_{i}} \,, \label{eq:assembled_vxc} 
    \end{align}
\end{subequations}
with the element matrices/vectors taking the form:
\begin{subequations}
    \begin{align}
[\kzero]^{(e, e')}_{i, j} =&  \iint \phi_{i}^{p_3(e)}(r) K_{0}(r,r';0)\phi_{j}^{p_3(e')}(r')\, drdr'  \,,\label{eq:element_chi0} \\
[\blambda_{\bxc}]^{(e)}_{i} =& \int \phi_{i}^{p_3(e)}(r)\Lambda_{xc}(r)\, dr \, . \label{eq:element_OEPRHS}
\end{align}
\end{subequations}
Above, the static response function $K_0$ is calculated using Eq.~\ref{eq:Chi0static}, while using the basis expansion of $\widetilde{R}_{nl}$ (Eq.~\ref{eq:Rtildediscretization}). Similarly, $\Lambda_{xc}$ is calculated  using Eq.~\ref{eq:OEPRHS}, while also using the basis expansion of $\widetilde{R}_{nl}$ (Eq.~\ref{eq:Rtildediscretization}).

In this work, we employ a polynomial mesh \cite{herman1963atomic, Uzulis_2022, gwasphericalatomshellgren} whose element size increases quadratically with radial distance. The quadratic variation, rather than the cubic one employed in previous RPA$-$OEP calculations \cite{gwasphericalatomshellgren}, is adopted because it provides improved numerical stability. The use of a polynomial mesh, instead of the exponential/logarithmic mesh commonly adopted in atomic-structure calculations with local/semilocal exchange--correlation functionals \cite{CERTIK2024109051, OEP-talman-5, GoodOEPaper, bhowmik2025spectral}, is motivated by the need to accurately resolve the increasingly delocalized unoccupied states required in RPA exchange--correlation calculations. In particular, enhanced resolution is required farther from the nucleus, which is more effectively achieved with a polynomial mesh than with an exponential mesh. By contrast, a uniform mesh would lead to an excessively fine discretization, particularly for high atomic numbers, owing to the all-electron nature of the calculations.

The nodes in each element are positioned according to the Chebyshev--Gauss--Lobatto scheme, and high-order Lagrange polynomials with \(C^{0}\)-continuity across elements are used as the basis functions \cite{PATERA1984468}. High-order approximations are critical in electronic structure calculations, since they enable the required accuracy to be attained without the use of excessively fine grids \cite{pask2005finite, SURYANARAYANA2010256, DFTFEVIKRAM}. However, employing high-order polynomials on a uniform grid can lead to the Runge phenomenon, which motivates the use of node distributions that provide increased mesh density near element boundaries. Basis functions with higher-order continuity across elements, such as \(C^{1}\)-continuous Hermite polynomials, are an attractive alternative that can reduce the number of basis functions required for a given accuracy, as demonstrated in previous atomic-structure calculations with local/semilocal exchange--correlation \cite{lehtolafem}. Given the one-dimensional nature of the present setting, the  choice of \(C^{0}\)-continuous Lagrange polynomials provides a good balance of simplicity and efficiency.

The spatial integrals are decomposed into element-wise contributions and evaluated using Gauss$-$Legendre quadrature. In spectral finite-element formulations for eigenproblems, Gauss--Lobatto quadrature is an attractive choice since it yields a diagonal overlap matrix, thereby allowing the generalized eigenvalue problem to be readily transformed into a standard eigenvalue problem that is significantly more efficient to solve \cite{DFTFEVIKRAM, CERTIK2024109051}. However, Gauss$-$Lobatto quadrature places quadrature nodes at the endpoints of the integration interval, which introduces difficulties associated with singular behavior in all-electron calculations such as those considered here. Moreover, the cost of solving the eigenvalue problem constitutes only a minor fraction of the total computational expense, which is instead dominated by the solution of the OEP equation for the local exchange--correlation potential. 

The OEP equation for the local exchange--correlation potential (Eq.~\ref{eq:FEOEP}) is decomposed into exchange and correlation contributions, which are solved separately. The solutions so obtained for the exchange and correlation potentials are shifted by a constant. The additive constant in the exchange potential is fixed using the HOMO condition \cite{OEP-singularity-2-ivanov}, which requires the expectation value of the local OEP exchange potential over the highest occupied orbital to equal that of the corresponding nonlocal exact-exchange operator. The additive constant in the correlation potential is determined by enforcing a zero boundary value at \( r = R_{\max} \).

We have developed an implementation of the above spectral finite-element framework  in \texttt{python}. The self-consistent solution is obtained by employing an outer iteration over the local exchange--correlation potential \( V_{xc} \), in which \( V_{xc} \) is held fixed while an inner self-consistent cycle is carried out with respect to the electron density. Direct solvers are used to compute the eigenproblem in Eq.~\ref{eq:GenEigenProblem} and to solve the linear systems in Eqs.~\ref{eq:FEPossion} and \ref{eq:FEOEP}. The inner self-consistent cycle is accelerated using the Periodic Pulay mixing scheme \cite{BANERJEE201631}. The implementation is parallelized over the frequencies in the quadrature rule, for which Gauss$-$Legendre quadrature is employed.


\section{Results and Discussion} \label{Sec:Results}
We now employ the  spectral finite-element framework to perform self-consistent RPA--OEP atomic structure calculations. In particular, we consider the following closed-shell atoms: helium (He), beryllium (Be), neon (Ne), magnesium (Mg), and argon (Ar), results for which  are available in the literature.  The numerical parameters are set as follows: the domain size is \(R_{\max} = 13\) bohr; the polynomial degrees are \(p_1 = 15\), \(p_2 = 31\), and \(p_3 = 4\); the spatial quadrature  order is \(55\); the numbers of finite-elements for He, Be, Ne, Mg, and Ar are \(N_e = 20\), \(30\), \(30\), \(40\), and \(60\), respectively; the maximum angular quantum  numbers are \(18\), \(18\), \(40\), \(40\), and \(45\), respectively; and the  frequency quadrature orders are \(30\), \(40\), \(60\), \(70\), and \(90\),  respectively. These choices ensure convergence of the quantities of interest, namely, ionization potential, HOMO--LUMO gap, total energy, and its difference from the  Hartree--Fock total energy, which can be interpreted as the correlation energy, to within $5\times 10^{-4}$~ha. 


\subsection{Accuracy}
We first assess the accuracy of the developed framework by comparing our results with those reported in Ref.~\onlinecite{gwasphericalatomshellgren}, which were obtained using a systematically improvable $\mathcal{C}^2$ cubic-spline basis and have served as reference data in the literature \cite{vacondiopaper, scRPAgorlingpaper}. Indeed, discrepancies of up to $0.15$~ha in the quantities of interest have been reported in the literature \cite{gwasphericalatomshellgren, scRPAgorlingpaper, OEP-correlation-RPA-1-hellgren-molecules}. These differences primarily stem from the need to achieve convergence with respect to a large number of numerical parameters, most notably the basis set and maximum angular momentum, as well as from the singular nature of the RPA--OEP equation, which makes the results particularly sensitive to the chosen numerical settings.

\begin{figure*}[htbp]
\subfigure[He]
{\includegraphics[keepaspectratio=true,width=0.325\textwidth]{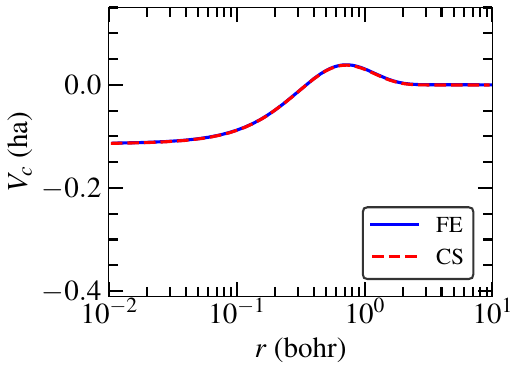}\label{He_RPA_correlation}}
\subfigure[Be]
{\includegraphics[keepaspectratio=true,width=0.325\textwidth]{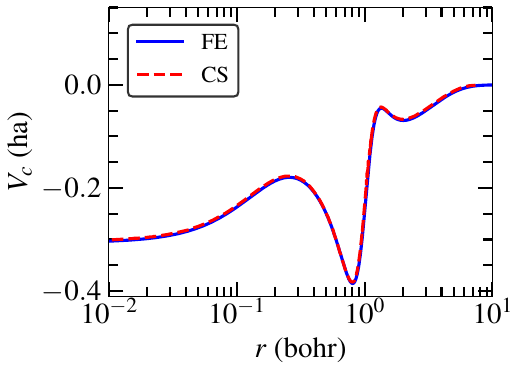}\label{Be_RPA_correlation}}
\subfigure[Ne]
{\includegraphics[keepaspectratio=true,width=0.325\textwidth]{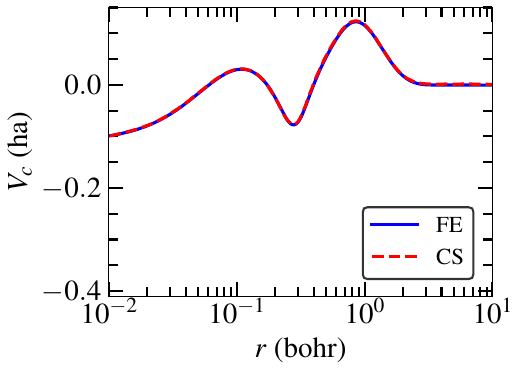}\label{Ne_RPA_correlation}}
\subfigure[Mg]
{\includegraphics[keepaspectratio=true,width=0.325\textwidth]{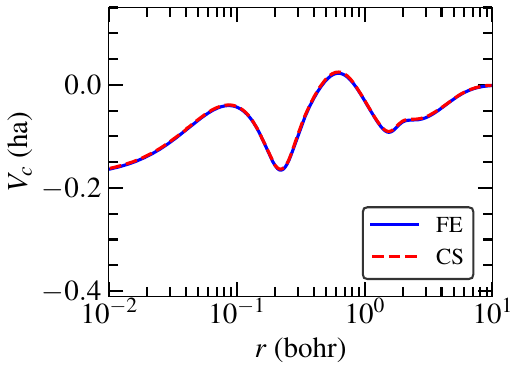}\label{Mg_RPA_correlation}}
\subfigure[Ar]
{\includegraphics[keepaspectratio=true,width=0.325\textwidth]{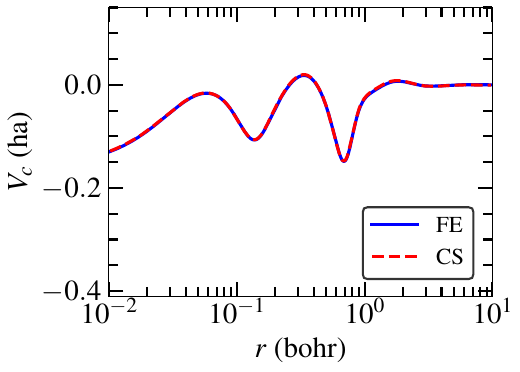}\label{Ar_RPA_correlation}}
\caption{\label{Fig:2RPA_Correlation_potentials} RPA--OEP correlation potentials obtained using the spectral finite-element (FE) framework,  along with the cubic-spline (CS) results reported in 
Ref.~\onlinecite{gwasphericalatomshellgren}, for (a) He, (b) Be, (c) Ne, (d) Mg, and (e) Ar.} 
\end{figure*}

In Fig.~\ref{Fig:2RPA_Correlation_potentials}, we compare the self-consistent RPA--OEP correlation  potentials obtained using the spectral finite-element framework with those  reported using the cubic-spline framework in Ref.~\onlinecite{gwasphericalatomshellgren}. Indeed, the RPA--OEP  exchange potential for all elements has not been provided in this reference, therefore our comparison is  restricted to the correlation component. We observe very good agreement between  the correlation potential curves, being practically indistinguishable for all chemical elements considered. In particular, the relative Frobenius-norm differences in the correlation potential for He, Be, Ne, Mg, and Ar are \(0.008\), \(0.028\), \(0.023\), \(0.017\), and \(0.016\), respectively.

In Table~\ref{Table:2}, we compare the quantities of interest, namely the ionization potential, HOMO--LUMO gap,  RPA--OEP total energy, and the difference between the RPA--OEP and Hartree--Fock total energies, obtained using the spectral finite-element framework, with the corresponding values reported for the cubic-spline framework in Ref.~\onlinecite{gwasphericalatomshellgren}. We observe that there is very good agreement, with the ionization potential differing by at most \(0.002\)~ha for Be/Mg/Ar, the HOMO--LUMO gap differing by at most \(0.001\)~ha for Ne, the difference between the RPA--OEP and Hartree-Fock total energies differing by at most \(0.011\)~ha for Ar, and the total energy differing by at most \(0.011\)~ha for Ar. Indeed, the differences increase with atomic number, with the largest discrepancies observed for Ar. This trend can likely be attributed to the aforementioned sensitivity to numerical parameters. In particular, the slower convergence of the quantities with respect to the maximum angular quantum number, details of which are not reported in the cited reference, is a possible source of the observed differences. Notably, the complete-basis-set--extrapolated total energies reported in Ref.~\onlinecite{scRPAgorlingpaper} differ from those obtained here by at most 0.002~ha for Ar. In addition to such extrapolation techniques, the convergence with discretization may be significantly accelerated by employing the Sternheimer equation to compute the orbital perturbations on a substantially finer grid, from which the density response function is evaluated, as explored for non-self-consistent RPA calculations \cite{nscRPA_sternheimer_atoms, nscRPA_sternheimer_molecules, nscRPA_delta_FE_molecules}.

\begin{table}[htbp]
\caption{\label{Table:2} RPA--OEP ionization potential (IP), HOMO-LUMO gap (Gap), total energy ($E$), and its difference from Hartree-Fock total energy ($E - E^{HF}$), obtained using the spectral finite-element (FE) framework, along with the cubic-spline (CS) results reported in  Ref.~\onlinecite{gwasphericalatomshellgren}. All quantities are reported in hartree.} 
\begin{ruledtabular}
\begin{tabular}{ccccccc}  
& & He&  Be&  Ne&  Mg& Ar \\ \hline
\multirow{2}{*}{$IP$} &
FE & -0.902  & -0.356 & -0.797& -0.299  &-0.592  \\
&CS & -0.902 & -0.354& -0.796 & -0.297 &-0.590  \\ \hline

\multirow{2}{*}{Gap} &
FE & -0.744 & -0.131& -0.608 & -0.128 & -0.431\\
&CS & -0.744 & -0.131 & -0.607 & -- & --\\ \hline

\multirow{2}{*}{${E}$} & 
 FE& -2.945 &  -14.754& -129.147 & -200.301& -527.919 \\
&CS & -2.945& -14.754 & -129.143 & -200.296 &-527.908 \\ \hline

\multirow{2}{*}{$E - E^{HF}$} & 
 FE & -0.084 & -0.181  & -0.599 & -0.687 & -1.102\\
&CS & -0.083 & -0.181 & -0.596 & -0.681 & -1.091 \\

\end{tabular}
\end{ruledtabular} 
\end{table}

The observed differences between the finite-element and cubic-spline results may, in part, reflect limitations of the cubic-spline basis in the present context. In particular, the global $\mathcal{C}^2$ continuity of cubic splines and their typical strong-form collocation increase inter-interval coupling and can lead to non-symmetric Hamiltonian matrices, with implications for both accuracy and efficiency. Finite-elements, by contrast, adopt a variational weak-form discretization that naturally preserves Hamiltonian symmetry, while employing strictly local $\mathcal{C}^0$ basis functions that more readily capture near-nuclear variations induced by the Coulomb singularity. Moreover, finite elements combine weaker inter-element coupling with a weak-form  treatment of derivatives, which promotes stable resolution of both occupied and unoccupied states, improves the conditioning of the resulting equations, and allows systematic refinement without reparameterization.  Finally, spectral finite-elements can achieve higher-order convergence under refinement, improving overall computational efficiency.

\subsection{Double-hybrid functional}
RPA is known to be relatively poor at describing short-range/local correlation effects, a consequence of neglecting the exchange--correlation kernel in the adiabatic-connection fluctuation–dissipation (ACFD) formalism. This has motivated the development of a number of variants, including RPA+ \cite{Perdewkurthrpaplus,perdewzidankurth} and its extensions \cite{RPA++perdew, double-hybrid-RPA-1, double-hybrid-RPA-3}. Here, we examine the fidelity of a one-parameter double-hybrid approximation \cite{double-hybrid-equation-1}. In its conventional form, this approximation employs M\o{}ller--Plesset second-order (MP2) correlation. In the present work, the MP2 correlation term is replaced by the RPA correlation: 
\begin{align}
    E_{xc} =& \,\alpha E_X + (1-\alpha) E_x^{\mathrm{GGA}} \nonumber \\
           & + \alpha^{3} E_c^{\mathrm{RPA}} + (1-\alpha^{3}) E_c^{\mathrm{GGA}} \,, \label{Eq:DH}
\end{align}
where $\alpha$ is a parameter, $E_X$ is the exact exchange energy, $E_x^{\mathrm{GGA}}$ and $E_c^{\mathrm{GGA}}$ are the exchange and correlation energies within the generalized gradient approximation (GGA), and $E_c^{\mathrm{RPA}}$ is the RPA correlation energy. Note that the linear dependence of exchange on $\alpha$ follows from the coordinate-scaling relation within the adiabatic-connection formalism. For correlation, coordinate scaling implies an $\alpha^{2}$ dependence; the additional factor of $\alpha$ arises when the density-scaled correlation is approximated by linearly interpolating between the weak-coupling (MP2) limit and a local/semilocal correlation \cite{double-hybrid-equation-1}.

\begin{figure*}[htbp]
\centering
\subfigure[He]
{\includegraphics[keepaspectratio=true,width=0.31\textwidth]{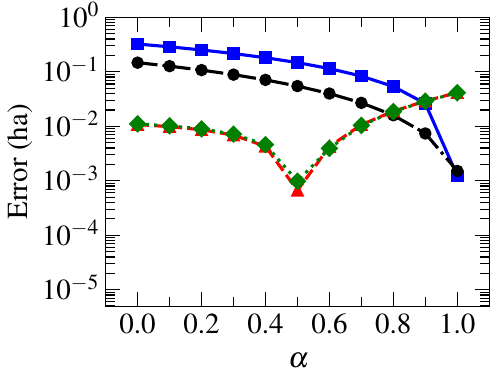}\label{He_double_hybrid}}
\subfigure[Be]
{\includegraphics[keepaspectratio=true,width=0.31\textwidth]{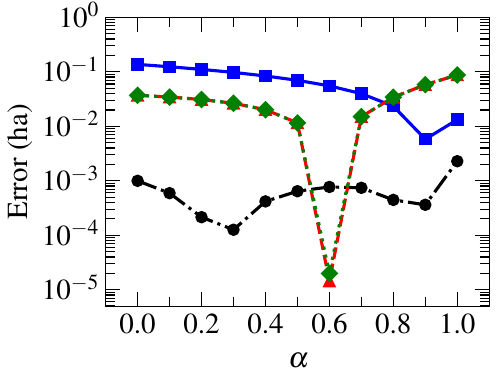}\label{Be_double_hybrid}}
\subfigure[Ne]
{\includegraphics[keepaspectratio=true,width=0.31\textwidth]{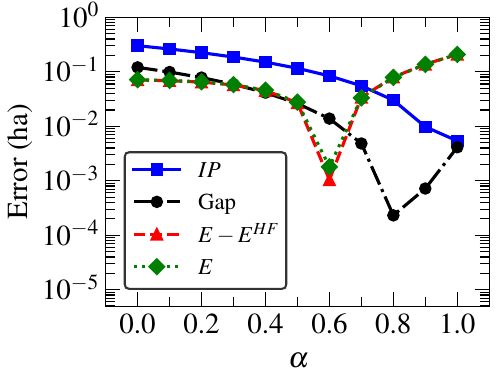}\label{Ne_double_hybrid}}
\subfigure[Mg]
{\includegraphics[keepaspectratio=true,width=0.31\textwidth]{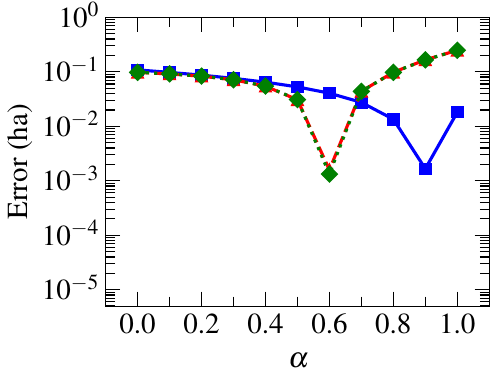}\label{Mg_double_hybrid}}
\subfigure[Ar]
{\includegraphics[keepaspectratio=true,width=0.31\textwidth]{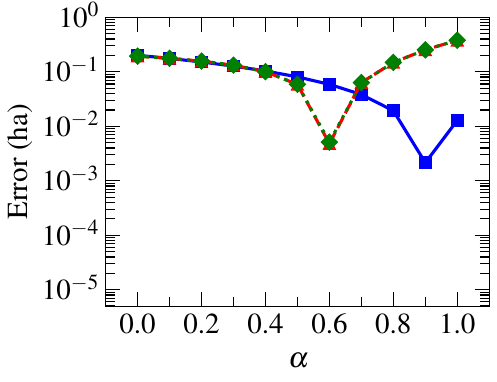}\label{Ar_double_hybrid}}
\caption{\label{Fig:3Double_hybrid} Variation of the error in ionization potential (IP), HOMO-LUMO gap (Gap), total energy ($E$), and its difference from the Hartree-Fock total energy ($E - E^{HF}$), with parameter $\alpha$ in the double-hybrid functional for (a) He, (b) Be, (c) Ne, (d) Mg, and (e) Ar. The reference values for IP, $E - E_{\mathrm{HF}}$, and $E$ are taken from full Configuration Interaction (CI) calculations \cite{chakravarthyberylliumandothers, chakravarthyhelium}, and the reference values for the Gap are obtained from inversion of quantum Monte Carlo (QMC) densities \cite{umrigargonze, gwasphericalatomshellgren}.} 
\end{figure*}

We perform self-consistent calculations within the OEP formalism for the double-hybrid exchange--correlation functional defined in Eq.~\ref{Eq:DH}, using the Perdew--Burke--Ernzerhof (PBE) \cite{PBE} functional for the GGA component. In Fig.~\ref{Fig:3Double_hybrid}, we present the dependence of the error in the quantities of interest, namely, ionization potential, HOMO-LUMO gap, total energy, and its difference from the Hartree-Fock total energy, on the parameter $\alpha$ appearing in the double hybrid functional. The reference values for the ionization potential, total energy, and difference from the Hartree-Fock total energy are obtained from full Configuration Interaction (CI) calculations \cite{chakravarthyberylliumandothers, chakravarthyhelium}, and the reference values for the HOMO--LUMO gap are obtained from inversion of quantum Monte Carlo (QMC) densities \cite{umrigargonze, gwasphericalatomshellgren}. We observe that the minimum error in the spectral properties, i.e., ionization potential and HOMO-LUMO gap, occur at values close to $\alpha = 1$, which corresponds to RPA--OEP, and the minimum error in energies, i.e., total energy and its difference from the total Hartree--Fock energy, occur at intermediate values of $\alpha$. This suggests that, while the single-parameter double-hybrid formalism can improve the total energies, it does so at the cost of generally poorer spectral properties.

In the present study, for simplicity, we focus on single-parameter double-hybrid functionals. Such functionals can be readily generalized by introducing separate parameters for the exchange and correlation components, potentially increasing accuracy. As an example, using a two-parameter double-hybrid  for helium with the exchange and correlation mixing parameters set to $0.9$ and $0.1$, respectively, yields errors of $0.002$, $0.020$, $0.003$, and $0.003$~ha in the ionization potential, HOMO--LUMO gap, total energy, and its difference from the Hartree--Fock total energy, respectively. These errors are comparable to the best results obtained with single-parameter double hybrids for the corresponding quantities, with minimum errors of \(0.001\), \(0.002\), \(0.001\), and \(0.0007\)~ha, respectively, attained at different values of the mixing parameter \(\alpha\).


\subsection{Machine-learned model}

The substantial computational and memory costs of RPA--OEP calculations, particularly in standard three-dimensional settings, motivate the development of lower-rung exchange--correlation functionals that provide comparable accuracy at substantially reduced cost. Here, we explore the feasibility of constructing a machine-learned model for the RPA--OEP exchange--correlation potential at the GGA level, i.e., exchange--correlation potential depends on the density and its gradient, calculations for which will be orders of magnitude faster than RPA--OEP in three-dimensional settings.

The  RPA--OEP exchange--correlation potential is modeled as:
\begin{align} \label{Eq:MLModel}
V_{xc}[\bx] = \sum_{t=1}^{N_t} w_t \mathcal{K} (\bx, \bx_t) \,,
\end{align}
where $V_{xc}[\bx]$ is the exchange--correlation potential corresponding to the descriptor $\bx$, $w_t$ are the model weights, and $\mathcal{K} (\bx, \bx_t)$ is a kernel that measures the similarity of the descriptor vectors $\bx$ and $\bx_t$, the latter corresponding to the data in the training dataset, of which there are $N_t$ instances. The exchange--correlation potential is standardized using the mean and standard deviation of the corresponding local density approximation (LDA) exchange--correlation potential. In this work, the Gaussian kernel is chosen to measure the similarity between the descriptors:
\begin{align}
    \mathcal{K}(\bx,\bx_t) = \exp\!\Bigg(-\frac{\big\lVert \bx - \bx_t \big\rVert_W^{2}}{2\sigma^2}\Bigg)\,, \label{eq:KRR_maternkernel}
\end{align}
where the weighted norm is defined as:
\begin{align}
    \big\lVert \bx - \bx_t \big\rVert_W
    = \sqrt{(\bx - \bx_t)^{\mathrm{T}} \mathbf{W} (\bx - \bx_t)} \, .
\end{align}
with $\mathbf{W}$ being a diagonal matrix of weights normalized such that their sum equals unity, i.e., $\mathbf{W}$ has unit trace. The descriptor is chosen to be:
\begin{align}
\bx = \begin{bmatrix}
    \rho, \, \frac{|\nabla \rho|}{\rho^{4/3}},  \, V_{xc}^{LDA}
\end{bmatrix} ^{\rm T}  \,,
\end{align}
where $\rho$ is the electron density,  $|\nabla \rho|/\rho^{4/3}$ is the reduced gradient,  and $V_{xc}^{LDA}$ is the LDA exchange-correlation potential.  Indeed, the reduced gradient is employed in place of the raw gradient, as commonly done in machine-learned exchange--correlation models \cite{Nagai2019CompletingDF, MLreducedgradientlaplacian, realspacesecondorderML, Reducedgradient3, MLvikram}, to ensure that the resulting exchange--correlation satisfies the scale-invariance property of the exact theory.  Each component of the descriptor is standardized by using the mean and standard deviation of the data for the associated chemical element.  The model weights are determined using linear regression, details for which can be found elsewhere \cite{deltalearning1, MLFFwarmdensematter, bishopML}.

The self-consistent data for the five chemical elements is interpolated using cubic splines onto a composite radial grid  comprising 1000 exponentially spaced points up to an electron density of \(0.01~\text{bohr}^{-3}\), followed by 1000 uniformly spaced points beyond this threshold. The interpolated data is then restricted to the radial interval \( r = 10^{-4} \) to \( 10 \)~bohr. The hyperparameters consist of $\sigma = 3$ and a regularization parameter for the linear regression set to $10^{-11}$. For prediction, the value of the RPA--OEP exchange--correlation potential at radial distances larger than 10 bohr is set to the value at \(r = 10\) bohr. A leave-one-out strategy is employed in which data from four of the five atoms are used for training, while the exchange--correlation potential for the remaining atom is predicted, with each atom in turn serving as the test case. The weights in $\mathbf{W}$ are optimized using the elitist micro genetic algorithm, which minimizes the testing error across the five atoms and yields $[\mathbf{W}]_{11}=0.7$, $[\mathbf{W}]_{22}=0.1$,  and $[\mathbf{W}]_{33}=0.2$. These values indicate that the LDA exchange--correlation potential constitutes an important component of the descriptor. Indeed, excluding this contribution leads to a significant increase in the error, particularly for the properties of interest.

\begin{figure*}[htbp]
\centering
\subfigure[He]
{\includegraphics[keepaspectratio=true,width=0.31\textwidth]{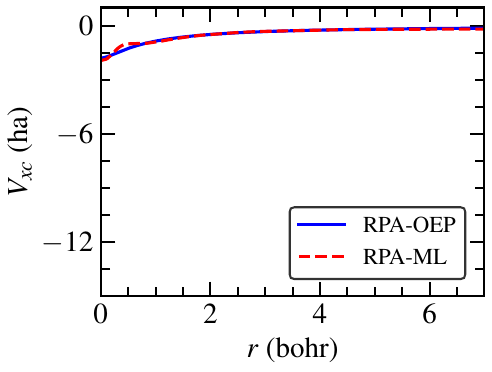}\label{He_ML_potential}}
\subfigure[Be]
{\includegraphics[keepaspectratio=true,width=0.31\textwidth]{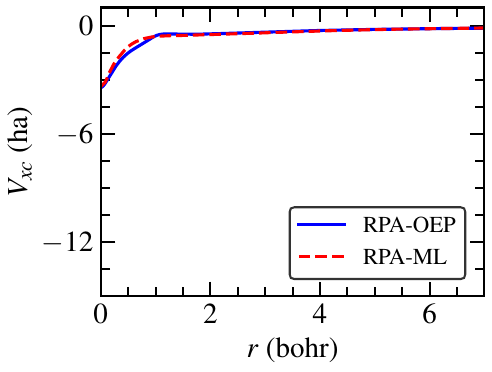}\label{Be_ML_potential}}
\subfigure[Ne]
{\includegraphics[keepaspectratio=true,width=0.31\textwidth]{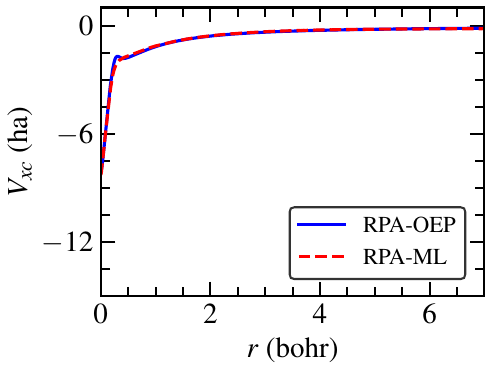}\label{Ne_ML_potential}}
\subfigure[Mg]
{\includegraphics[keepaspectratio=true,width=0.31\textwidth]{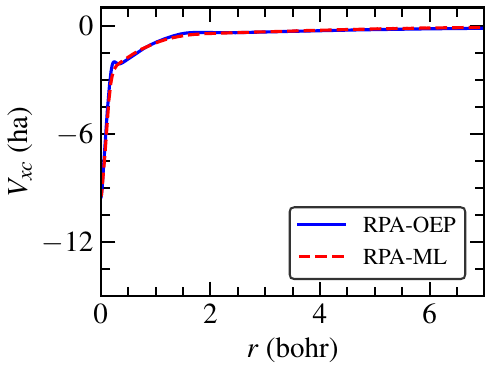}\label{Mg_ML_potential}}
\subfigure[Ar]
{\includegraphics[keepaspectratio=true,width=0.31\textwidth]{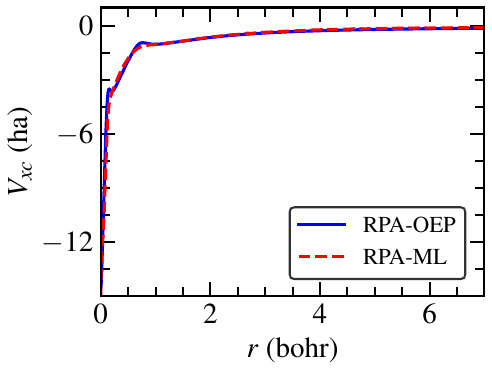}\label{Ar_ML_potential}}
\caption{\label{Fig:4ML_potential} Comparison of the RPA--OEP exchange--correlation potential with that predicted from the machine-learned model (RPA--ML) for (a) He, (b) Be, (c) Ne, (d) Mg, and (e) Ar. In each case, the model is trained on the remaining atoms and used to predict the target atom.} 
\end{figure*}

In Fig.~\ref{Fig:4ML_potential}, we compare the RPA--OEP exchange--correlation potential with that predicted by the machine-learned model. We find reasonably good agreement, especially given the limited amount of training data. In particular, the relative Frobenius-norm differences in the correlation potential for He, Be, Ne, Mg, and Ar are 0.113, 0.092, 0.041, 0.058, and 0.092, respectively. Indeed, these values are comparable to the aforementioned differences in the RPA--OEP exchange--correlation potential obtained here relative to the cubic-spline framework reported in Ref.~\onlinecite{gwasphericalatomshellgren}.

In Table~\ref{Table:ML}, we report the differences in the quantities of interest, namely the ionization potential, HOMO--LUMO gap, and total energy, between the machine-learned model predictions and the RPA--OEP reference results. For the machine-learned model results, we do a self-consistent solution, while holding the exchange--correlation potential fixed. For purposes of comparison, we also show the PBE \cite{PBE} and  r\textsuperscript{2}SCAN \cite{r2SCAN} exchange--correlation results. We observe that in the spectral properties, i.e., ionization potential and HOMO-LUMO gap, the machine-learned model results are very close to the RPA--OEP results, much closer than the PBE and r\textsuperscript{2}SCAN results. Though the energies from the machine-learned model are also in good agreement, they are further off than the PBE and r\textsuperscript{2}SCAN results. This can partly be explained by the fact that the fitting is done to the RPA--OEP exchange--correlation potential, which does not directly translate to the energies.  Indeed, the RPA-ML exchange--correlation potential does not minimize the RPA energy functional. Interestingly, evaluating the PBE energy functional at the RPA-ML ground state leads to a significant improvement in the energy values.

\begin{table}[htbp!]
\caption{\label{Table:ML}Differences relative to the RPA--OEP reference values for the ionization potential (IP), HOMO--LUMO gap (Gap), and total energy ($E$), obtained using the machine-learned model (RPA-ML). The RPA-ML results correspond to the self-consistent solution, while holding the exchange--correlation potential fixed. The PBE and r\textsuperscript{2}SCAN exchange-correlation results are also shown. All quantities are reported in hartree.} 
\begin{ruledtabular}
\begin{tabular}{ccccccc}  
& & He&  Be&  Ne&  Mg& Ar \\ \hline
\multirow{3}{*}{$IP$} &
RPA-ML& 0.024 & -0.033 & 0.019 & 0.011 & 0.020  \\
&PBE & 0.322 & 0.150 & 0.307 & 0.127  & 0.214 \\
&r\textsuperscript{2}SCAN & 0.294 & 0.145 & 0.283 & 0.122 & 0.198 \\
\hline
\multirow{3}{*}{Gap} &
RPA-ML  & 0.045 & -0.007 & 0.015 & -0.013 & -0.018 \\
&PBE & 0.146 & -0.001 & 0.117  & 0.004 & 0.059 \\
&r\textsuperscript{2}SCAN & 0.074 & -0.015  & 0.074 & -0.006 & 0.019 \\ \hline 
\multirow{3}{*}{${E}$} & 
RPA-ML& 0.004 & 0.002 & 0.003  & 0.008 & 0.019  \\
&PBE & 0.001 & 0.002 & 0.002  & 0.002 & 0.003 \\
&r\textsuperscript{2}SCAN & 0.002  & 0.008 & 0.014 & 0.015 &0.024\\ 
\end{tabular}
\end{ruledtabular} 
\end{table}

In the present study, we consider only five atoms, and therefore the data available for training and testing the machine-learned model is limited. In view of this, we employ a leave-one-out strategy, in which the model is trained on four atoms and evaluated on the remaining atom. An important direction for future work is the development of a single unified model trained on data drawn from all chemical elements, including charged atoms and fictitious atoms with fractional occupations; such data are designed to mimic atomic environments encountered in molecular and bulk-like systems. Leveraging these larger and more diverse data sets would enable the construction of machine-learned exchange--correlation models that are broadly transferable and potentially applicable to general 3D systems.

The model developed here bears similarities to that of Ref.~\onlinecite{MLRPAKresse}, which likewise employs a kernel-based approach and linear regression for the weights. However, they are a number of key differences. First, the model in Ref.~\onlinecite{MLRPAKresse} is constructed at the level of the energy, from which the corresponding potential is subsequently derived; this leads to a formulation that differs from the present approach, where the exchange--correlation potential is fitted directly. As a consequence, a limitation of the current work is that the evaluation of the energy remains computationally expensive. Second, Ref.~\onlinecite{MLRPAKresse} employs a loss function that incorporates both the energy and the potential, while including the electron density as a weighting function. Third and finally, while the descriptors used in the present work include the LDA exchange--correlation potential, Ref.~\onlinecite{MLRPAKresse} restricts the descriptor to include only the density and its gradient.


\section{Concluding Remarks} \label{Sec:Conclusions}

We developed a spectral finite-element formulation of the OEP method for atomic structure calculations with the RPA exchange-correlation. The framework employs a polynomial mesh with element nodes distributed according to the Chebyshev--Gauss--Lobatto scheme, high-order $\mathcal{C}^0$-continuous Lagrange polynomial basis functions, and Gauss--Legendre quadrature for spatial integration. Different polynomial degrees were adopted for the orbitals, Hartree potential, and RPA--OEP exchange--correlation potential. Using He, Be, Ne, Mg, and Ar as representative examples, we  established the accuracy of the developed framework. In addition, we examined the fidelity of one-parameter double-hybrid functionals constructed with RPA correlation, finding that while they improve the energetics, they tend to degrade spectral properties. Finally, we developed a machine-learned model  for the RPA--OEP exchange--correlation potential at the GGA level, based on kernel methods and linear regression, while employing a descriptor constructed from the electron density, its reduced gradient, and the LDA potential. Despite the limited size of the training dataset, the model exhibited good predictive performance.

Future work includes extending the present OEP framework to other Green’s function--based exchange-correlation. In addition, generalizing the framework to charged atoms and fractional occupations, thereby mimicking atomic environments encountered in molecules and bulk-like conditions, would enable the systematic generation of larger and more diverse data sets, providing a foundation for the development of more accurate machine-learned models for the exchange-correlation.

\begin{acknowledgments}
The authors gratefully acknowledge the support of the U.S. Department of Energy, Office of Science under grant DE-SC0023445. This research was also supported by the supercomputing infrastructure provided by Partnership for an Advanced Computing Environment (PACE) through its Hive (U.S. National Science Foundation through grant MRI1828187) and Phoenix clusters at Georgia Institute of Technology, Atlanta, Georgia. The authors acknowledge insightful discussions with Andrew J. Medford and John E. Pask.
\end{acknowledgments}
\section*{Data Availability Statement}
The data that support the findings of this study are available within the article and from the corresponding author upon reasonable request.
\section*{Author Declarations}
The authors have no conflicts to disclose.
\section*{References}
\vspace{-5mm}

\begin{thebibliography}{84}%
\makeatletter
\providecommand \@ifxundefined [1]{%
 \@ifx{#1\undefined}
}%
\providecommand \@ifnum [1]{%
 \ifnum #1\expandafter \@firstoftwo
 \else \expandafter \@secondoftwo
 \fi
}%
\providecommand \@ifx [1]{%
 \ifx #1\expandafter \@firstoftwo
 \else \expandafter \@secondoftwo
 \fi
}%
\providecommand \natexlab [1]{#1}%
\providecommand \enquote  [1]{``#1''}%
\providecommand \bibnamefont  [1]{#1}%
\providecommand \bibfnamefont [1]{#1}%
\providecommand \citenamefont [1]{#1}%
\providecommand \href@noop [0]{\@secondoftwo}%
\providecommand \href [0]{\begingroup \@sanitize@url \@href}%
\providecommand \@href[1]{\@@startlink{#1}\@@href}%
\providecommand \@@href[1]{\endgroup#1\@@endlink}%
\providecommand \@sanitize@url [0]{\catcode `\\12\catcode `\$12\catcode
  `\&12\catcode `\#12\catcode `\^12\catcode `\_12\catcode `\%12\relax}%
\providecommand \@@startlink[1]{}%
\providecommand \@@endlink[0]{}%
\providecommand \url  [0]{\begingroup\@sanitize@url \@url }%
\providecommand \@url [1]{\endgroup\@href {#1}{\urlprefix }}%
\providecommand \urlprefix  [0]{URL }%
\providecommand \Eprint [0]{\href }%
\providecommand \doibase [0]{http://dx.doi.org/}%
\providecommand \selectlanguage [0]{\@gobble}%
\providecommand \bibinfo  [0]{\@secondoftwo}%
\providecommand \bibfield  [0]{\@secondoftwo}%
\providecommand \translation [1]{[#1]}%
\providecommand \BibitemOpen [0]{}%
\providecommand \bibitemStop [0]{}%
\providecommand \bibitemNoStop [0]{.\EOS\space}%
\providecommand \EOS [0]{\spacefactor3000\relax}%
\providecommand \BibitemShut  [1]{\csname bibitem#1\endcsname}%
\let\auto@bib@innerbib\@empty
\bibitem [{\citenamefont {Kohn}\ and\ \citenamefont
  {Sham}(1965)}]{Kohn-shampaper}%
  \BibitemOpen
  \bibfield  {author} {\bibinfo {author} {\bibfnamefont {W.}~\bibnamefont
  {Kohn}}\ and\ \bibinfo {author} {\bibfnamefont {L.~J.}\ \bibnamefont
  {Sham}},\ }\href@noop {} {\bibfield  {journal} {\bibinfo  {journal} {Phys.
  Rev.}\ }\textbf {\bibinfo {volume} {140}},\ \bibinfo {pages} {A1133}
  (\bibinfo {year} {1965})}\BibitemShut {NoStop}%
\bibitem [{\citenamefont {Hohenberg}\ and\ \citenamefont
  {Kohn}(1964)}]{Hohenberg-Kohnpaper}%
  \BibitemOpen
  \bibfield  {author} {\bibinfo {author} {\bibfnamefont {P.}~\bibnamefont
  {Hohenberg}}\ and\ \bibinfo {author} {\bibfnamefont {W.}~\bibnamefont
  {Kohn}},\ }\href@noop {} {\bibfield  {journal} {\bibinfo  {journal} {Phys.
  Rev.}\ }\textbf {\bibinfo {volume} {136}},\ \bibinfo {pages} {B864} (\bibinfo
  {year} {1964})}\BibitemShut {NoStop}%
\bibitem [{\citenamefont {Perdew}\ and\ \citenamefont
  {Schmidt}(2001)}]{jacobladderperdew}%
  \BibitemOpen
  \bibfield  {author} {\bibinfo {author} {\bibfnamefont {J.~P.}\ \bibnamefont
  {Perdew}}\ and\ \bibinfo {author} {\bibfnamefont {K.}~\bibnamefont
  {Schmidt}},\ }\href@noop {} {\bibfield  {journal} {\bibinfo  {journal} {AIP
  Conf. Proc.}\ }\textbf {\bibinfo {volume} {577}},\ \bibinfo {pages} {1}
  (\bibinfo {year} {2001})}\BibitemShut {NoStop}%
\bibitem [{\citenamefont {Langreth}\ and\ \citenamefont
  {Perdew}(1977)}]{ACFDperdew}%
  \BibitemOpen
  \bibfield  {author} {\bibinfo {author} {\bibfnamefont {D.~C.}\ \bibnamefont
  {Langreth}}\ and\ \bibinfo {author} {\bibfnamefont {J.~P.}\ \bibnamefont
  {Perdew}},\ }\href@noop {} {\bibfield  {journal} {\bibinfo  {journal} {Phys.
  Rev. B}\ }\textbf {\bibinfo {volume} {15}},\ \bibinfo {pages} {2884}
  (\bibinfo {year} {1977})}\BibitemShut {NoStop}%
\bibitem [{\citenamefont {Ren}\ \emph {et~al.}(2012)\citenamefont {Ren},
  \citenamefont {Rinke}, \citenamefont {Joas},\ and\ \citenamefont
  {Scheffler}}]{ren2012random}%
  \BibitemOpen
  \bibfield  {author} {\bibinfo {author} {\bibfnamefont {X.}~\bibnamefont
  {Ren}}, \bibinfo {author} {\bibfnamefont {P.}~\bibnamefont {Rinke}}, \bibinfo
  {author} {\bibfnamefont {C.}~\bibnamefont {Joas}}, \ and\ \bibinfo {author}
  {\bibfnamefont {M.}~\bibnamefont {Scheffler}},\ }\href@noop {} {\bibfield
  {journal} {\bibinfo  {journal} {J. Mater. Sci.}\ }\textbf {\bibinfo {volume}
  {47}},\ \bibinfo {pages} {7447} (\bibinfo {year} {2012})}\BibitemShut
  {NoStop}%
\bibitem [{\citenamefont {Eshuis}, \citenamefont {Bates},\ and\ \citenamefont
  {Furche}(2012)}]{eshuis2012electron}%
  \BibitemOpen
  \bibfield  {author} {\bibinfo {author} {\bibfnamefont {H.}~\bibnamefont
  {Eshuis}}, \bibinfo {author} {\bibfnamefont {J.~E.}\ \bibnamefont {Bates}}, \
  and\ \bibinfo {author} {\bibfnamefont {F.}~\bibnamefont {Furche}},\
  }\href@noop {} {\bibfield  {journal} {\bibinfo  {journal} {Theor. Chem.
  Acc.}\ }\textbf {\bibinfo {volume} {131}},\ \bibinfo {pages} {1084} (\bibinfo
  {year} {2012})}\BibitemShut {NoStop}%
\bibitem [{\citenamefont {Ren}, \citenamefont {Rinke},\ and\ \citenamefont
  {Scheffler}(2009)}]{RenRPA3}%
  \BibitemOpen
  \bibfield  {author} {\bibinfo {author} {\bibfnamefont {X.}~\bibnamefont
  {Ren}}, \bibinfo {author} {\bibfnamefont {P.}~\bibnamefont {Rinke}}, \ and\
  \bibinfo {author} {\bibfnamefont {M.}~\bibnamefont {Scheffler}},\ }\href@noop
  {} {\bibfield  {journal} {\bibinfo  {journal} {Phys. Rev. B}\ }\textbf
  {\bibinfo {volume} {80}},\ \bibinfo {pages} {045402} (\bibinfo {year}
  {2009})}\BibitemShut {NoStop}%
\bibitem [{\citenamefont {Harl}, \citenamefont {Schimka},\ and\ \citenamefont
  {Kresse}(2010)}]{KresseRPAlattice}%
  \BibitemOpen
  \bibfield  {author} {\bibinfo {author} {\bibfnamefont {J.}~\bibnamefont
  {Harl}}, \bibinfo {author} {\bibfnamefont {L.}~\bibnamefont {Schimka}}, \
  and\ \bibinfo {author} {\bibfnamefont {G.}~\bibnamefont {Kresse}},\
  }\href@noop {} {\bibfield  {journal} {\bibinfo  {journal} {Phys. Rev. B}\
  }\textbf {\bibinfo {volume} {81}},\ \bibinfo {pages} {115126} (\bibinfo
  {year} {2010})}\BibitemShut {NoStop}%
\bibitem [{\citenamefont {Del~Ben}, \citenamefont {Hutter},\ and\ \citenamefont
  {VandeVondele}(2015)}]{Hutterliquidwater}%
  \BibitemOpen
  \bibfield  {author} {\bibinfo {author} {\bibfnamefont {M.}~\bibnamefont
  {Del~Ben}}, \bibinfo {author} {\bibfnamefont {J.}~\bibnamefont {Hutter}}, \
  and\ \bibinfo {author} {\bibfnamefont {J.}~\bibnamefont {VandeVondele}},\
  }\href@noop {} {\bibfield  {journal} {\bibinfo  {journal} {J. Chem. Phys.}\
  }\textbf {\bibinfo {volume} {143}},\ \bibinfo {pages} {054506} (\bibinfo
  {year} {2015})}\BibitemShut {NoStop}%
\bibitem [{\citenamefont {Cui}, \citenamefont {Wu},\ and\ \citenamefont
  {Jiang}(2016)}]{JiangRPAstability}%
  \BibitemOpen
  \bibfield  {author} {\bibinfo {author} {\bibfnamefont {Z.-H.}\ \bibnamefont
  {Cui}}, \bibinfo {author} {\bibfnamefont {F.}~\bibnamefont {Wu}}, \ and\
  \bibinfo {author} {\bibfnamefont {H.}~\bibnamefont {Jiang}},\ }\href@noop {}
  {\bibfield  {journal} {\bibinfo  {journal} {Phys. Chem. Chem. Phys.}\
  }\textbf {\bibinfo {volume} {18}},\ \bibinfo {pages} {29914} (\bibinfo {year}
  {2016})}\BibitemShut {NoStop}%
\bibitem [{\citenamefont {Hermann}, \citenamefont {DiStasio},\ and\
  \citenamefont {Tkatchenko}(2017)}]{Tkatchenkovanderwaals}%
  \BibitemOpen
  \bibfield  {author} {\bibinfo {author} {\bibfnamefont {J.}~\bibnamefont
  {Hermann}}, \bibinfo {author} {\bibfnamefont {R.~A.~J.}\ \bibnamefont
  {DiStasio}}, \ and\ \bibinfo {author} {\bibfnamefont {A.}~\bibnamefont
  {Tkatchenko}},\ }\href@noop {} {\bibfield  {journal} {\bibinfo  {journal}
  {Chem. Rev.}\ }\textbf {\bibinfo {volume} {117}},\ \bibinfo {pages} {4714}
  (\bibinfo {year} {2017})}\BibitemShut {NoStop}%
\bibitem [{\citenamefont {Schmidt}\ and\ \citenamefont
  {Thygesen}(2018)}]{ThygesenAdsorptionenergies}%
  \BibitemOpen
  \bibfield  {author} {\bibinfo {author} {\bibfnamefont {P.~S.}\ \bibnamefont
  {Schmidt}}\ and\ \bibinfo {author} {\bibfnamefont {K.~S.}\ \bibnamefont
  {Thygesen}},\ }\href@noop {} {\bibfield  {journal} {\bibinfo  {journal} {J.
  Phys. Chem. C}\ }\textbf {\bibinfo {volume} {122}},\ \bibinfo {pages} {4381}
  (\bibinfo {year} {2018})}\BibitemShut {NoStop}%
\bibitem [{\citenamefont {Oudot}\ and\ \citenamefont
  {Doblhoff-Dier}(2024)}]{DierRPA}%
  \BibitemOpen
  \bibfield  {author} {\bibinfo {author} {\bibfnamefont {B.}~\bibnamefont
  {Oudot}}\ and\ \bibinfo {author} {\bibfnamefont {K.}~\bibnamefont
  {Doblhoff-Dier}},\ }\href@noop {} {\bibfield  {journal} {\bibinfo  {journal}
  {J. Chem. Phys.}\ }\textbf {\bibinfo {volume} {161}},\ \bibinfo {pages}
  {054708} (\bibinfo {year} {2024})}\BibitemShut {NoStop}%
\bibitem [{\citenamefont {Pitts}, \citenamefont {Contant},\ and\ \citenamefont
  {Hellgren}(2025)}]{RPA-SELF-CONSISTENT-PBE0-PITTS-PSEUDOPOTENTIAL-SOLIDS}%
  \BibitemOpen
  \bibfield  {author} {\bibinfo {author} {\bibfnamefont {T.}~\bibnamefont
  {Pitts}}, \bibinfo {author} {\bibfnamefont {D.}~\bibnamefont {Contant}}, \
  and\ \bibinfo {author} {\bibfnamefont {M.}~\bibnamefont {Hellgren}},\
  }\href@noop {} {\bibfield  {journal} {\bibinfo  {journal} {Phys. Rev. B}\
  }\textbf {\bibinfo {volume} {112}},\ \bibinfo {pages} {085137} (\bibinfo
  {year} {2025})}\BibitemShut {NoStop}%
\bibitem [{\citenamefont {Ramberger}, \citenamefont {Sch\"afer},\ and\
  \citenamefont {Kresse}(2017)}]{KresseRPAForces}%
  \BibitemOpen
  \bibfield  {author} {\bibinfo {author} {\bibfnamefont {B.}~\bibnamefont
  {Ramberger}}, \bibinfo {author} {\bibfnamefont {T.}~\bibnamefont
  {Sch\"afer}}, \ and\ \bibinfo {author} {\bibfnamefont {G.}~\bibnamefont
  {Kresse}},\ }\href@noop {} {\bibfield  {journal} {\bibinfo  {journal} {Phys.
  Rev. Lett.}\ }\textbf {\bibinfo {volume} {118}},\ \bibinfo {pages} {106403}
  (\bibinfo {year} {2017})}\BibitemShut {NoStop}%
\bibitem [{\citenamefont {Zhang}\ \emph {et~al.}(2025)\citenamefont {Zhang},
  \citenamefont {Shah}, \citenamefont {Pask}, \citenamefont {Chow},\ and\
  \citenamefont {Suryanarayana}}]{boqinRPA}%
  \BibitemOpen
  \bibfield  {author} {\bibinfo {author} {\bibfnamefont {B.}~\bibnamefont
  {Zhang}}, \bibinfo {author} {\bibfnamefont {S.}~\bibnamefont {Shah}},
  \bibinfo {author} {\bibfnamefont {J.~E.}\ \bibnamefont {Pask}}, \bibinfo
  {author} {\bibfnamefont {E.}~\bibnamefont {Chow}}, \ and\ \bibinfo {author}
  {\bibfnamefont {P.}~\bibnamefont {Suryanarayana}},\ }\href@noop {} {\bibfield
   {journal} {\bibinfo  {journal} {J. Chem. Theory Comput.}\ }\textbf {\bibinfo
  {volume} {21}},\ \bibinfo {pages} {6023} (\bibinfo {year}
  {2025})}\BibitemShut {NoStop}%
\bibitem [{\citenamefont {Shah}\ \emph {et~al.}(2024)\citenamefont {Shah},
  \citenamefont {Zhang}, \citenamefont {Huang}, \citenamefont {Pask},
  \citenamefont {Suryanarayana},\ and\ \citenamefont {Chow}}]{ShikharRPA}%
  \BibitemOpen
  \bibfield  {author} {\bibinfo {author} {\bibfnamefont {S.}~\bibnamefont
  {Shah}}, \bibinfo {author} {\bibfnamefont {B.}~\bibnamefont {Zhang}},
  \bibinfo {author} {\bibfnamefont {H.}~\bibnamefont {Huang}}, \bibinfo
  {author} {\bibfnamefont {J.~E.}\ \bibnamefont {Pask}}, \bibinfo {author}
  {\bibfnamefont {P.}~\bibnamefont {Suryanarayana}}, \ and\ \bibinfo {author}
  {\bibfnamefont {E.}~\bibnamefont {Chow}},\ }in\ \href@noop {} {\emph
  {\bibinfo {booktitle} {SC24: International Conference for High Performance
  Computing, Networking, Storage and Analysis}}}\ (\bibinfo {year} {2024})\
  pp.\ \bibinfo {pages} {1--15}\BibitemShut {NoStop}%
\bibitem [{\citenamefont {Talman}\ and\ \citenamefont
  {Shadwick}(1976)}]{oepjdtalman}%
  \BibitemOpen
  \bibfield  {author} {\bibinfo {author} {\bibfnamefont {J.~D.}\ \bibnamefont
  {Talman}}\ and\ \bibinfo {author} {\bibfnamefont {W.~F.}\ \bibnamefont
  {Shadwick}},\ }\href@noop {} {\bibfield  {journal} {\bibinfo  {journal}
  {Phys. Rev. A}\ }\textbf {\bibinfo {volume} {14}},\ \bibinfo {pages} {36}
  (\bibinfo {year} {1976})}\BibitemShut {NoStop}%
\bibitem [{\citenamefont {Sham}\ and\ \citenamefont
  {Schl\"uter}(1983)}]{Sham-Schluter-equation-OEP}%
  \BibitemOpen
  \bibfield  {author} {\bibinfo {author} {\bibfnamefont {L.~J.}\ \bibnamefont
  {Sham}}\ and\ \bibinfo {author} {\bibfnamefont {M.}~\bibnamefont
  {Schl\"uter}},\ }\href@noop {} {\bibfield  {journal} {\bibinfo  {journal}
  {Phys. Rev. Lett.}\ }\textbf {\bibinfo {volume} {51}},\ \bibinfo {pages}
  {1888} (\bibinfo {year} {1983})}\BibitemShut {NoStop}%
\bibitem [{\citenamefont {Bhowmik}\ \emph {et~al.}(2025)\citenamefont
  {Bhowmik}, \citenamefont {Pask}, \citenamefont {Medford},\ and\ \citenamefont
  {Suryanarayana}}]{bhowmik2025spectral}%
  \BibitemOpen
  \bibfield  {author} {\bibinfo {author} {\bibfnamefont {S.}~\bibnamefont
  {Bhowmik}}, \bibinfo {author} {\bibfnamefont {J.~E.}\ \bibnamefont {Pask}},
  \bibinfo {author} {\bibfnamefont {A.~J.}\ \bibnamefont {Medford}}, \ and\
  \bibinfo {author} {\bibfnamefont {P.}~\bibnamefont {Suryanarayana}},\
  }\href@noop {} {\bibfield  {journal} {\bibinfo  {journal} {Comput. Phys.
  Commun.}\ }\textbf {\bibinfo {volume} {308}},\ \bibinfo {pages} {109448}
  (\bibinfo {year} {2025})}\BibitemShut {NoStop}%
\bibitem [{\citenamefont {Čertík}, \citenamefont {Pask},\ and\ \citenamefont
  {Vackář}(2013)}]{dftatom}%
  \BibitemOpen
  \bibfield  {author} {\bibinfo {author} {\bibfnamefont {O.}~\bibnamefont
  {Čertík}}, \bibinfo {author} {\bibfnamefont {J.~E.}\ \bibnamefont {Pask}},
  \ and\ \bibinfo {author} {\bibfnamefont {J.}~\bibnamefont {Vackář}},\
  }\href@noop {} {\bibfield  {journal} {\bibinfo  {journal} {Comput. Phys.
  Commun.}\ }\textbf {\bibinfo {volume} {184}},\ \bibinfo {pages} {1777}
  (\bibinfo {year} {2013})}\BibitemShut {NoStop}%
\bibitem [{\citenamefont {Čertík}\ \emph {et~al.}(2024)\citenamefont
  {Čertík}, \citenamefont {Pask}, \citenamefont {Fernando}, \citenamefont
  {Goswami}, \citenamefont {Sukumar}, \citenamefont {Collins}, \citenamefont
  {Manzini},\ and\ \citenamefont {Vackář}}]{CERTIK2024109051}%
  \BibitemOpen
  \bibfield  {author} {\bibinfo {author} {\bibfnamefont {O.}~\bibnamefont
  {Čertík}}, \bibinfo {author} {\bibfnamefont {J.~E.}\ \bibnamefont {Pask}},
  \bibinfo {author} {\bibfnamefont {I.}~\bibnamefont {Fernando}}, \bibinfo
  {author} {\bibfnamefont {R.}~\bibnamefont {Goswami}}, \bibinfo {author}
  {\bibfnamefont {N.}~\bibnamefont {Sukumar}}, \bibinfo {author} {\bibfnamefont
  {L.~A.}\ \bibnamefont {Collins}}, \bibinfo {author} {\bibfnamefont
  {G.}~\bibnamefont {Manzini}}, \ and\ \bibinfo {author} {\bibfnamefont
  {J.}~\bibnamefont {Vackář}},\ }\href@noop {} {\bibfield  {journal}
  {\bibinfo  {journal} {Comput. Phys. Commun.}\ }\textbf {\bibinfo {volume}
  {297}},\ \bibinfo {pages} {109051} (\bibinfo {year} {2024})}\BibitemShut
  {NoStop}%
\bibitem [{\citenamefont {Lehtola}(2023{\natexlab{a}})}]{lehtolafem}%
  \BibitemOpen
  \bibfield  {author} {\bibinfo {author} {\bibfnamefont {S.}~\bibnamefont
  {Lehtola}},\ }\href@noop {} {\bibfield  {journal} {\bibinfo  {journal} {J.
  Phys. Chem. A}\ }\textbf {\bibinfo {volume} {127}},\ \bibinfo {pages} {4180}
  (\bibinfo {year} {2023}{\natexlab{a}})}\BibitemShut {NoStop}%
\bibitem [{\citenamefont {Lehtola}(2019)}]{lehtolafem2}%
  \BibitemOpen
  \bibfield  {author} {\bibinfo {author} {\bibfnamefont {S.}~\bibnamefont
  {Lehtola}},\ }\href@noop {} {\bibfield  {journal} {\bibinfo  {journal} {Int.
  J. Quantum Chem.}\ }\textbf {\bibinfo {volume} {119}},\ \bibinfo {pages}
  {e25945} (\bibinfo {year} {2019})}\BibitemShut {NoStop}%
\bibitem [{\citenamefont {Lehtola}(2023{\natexlab{b}})}]{lehtolafem3}%
  \BibitemOpen
  \bibfield  {author} {\bibinfo {author} {\bibfnamefont {S.}~\bibnamefont
  {Lehtola}},\ }\href@noop {} {\bibfield  {journal} {\bibinfo  {journal} {J.
  Chem. Theory Comput.}\ }\textbf {\bibinfo {volume} {19}},\ \bibinfo {pages}
  {2502} (\bibinfo {year} {2023}{\natexlab{b}})}\BibitemShut {NoStop}%
\bibitem [{\citenamefont {Lehtola}(2020)}]{lehtolafem4}%
  \BibitemOpen
  \bibfield  {author} {\bibinfo {author} {\bibfnamefont {S.}~\bibnamefont
  {Lehtola}},\ }\href@noop {} {\bibfield  {journal} {\bibinfo  {journal} {Phys.
  Rev. A}\ }\textbf {\bibinfo {volume} {101}},\ \bibinfo {pages} {012516}
  (\bibinfo {year} {2020})}\BibitemShut {NoStop}%
\bibitem [{\citenamefont {Romanowski}(2007)}]{Romanowski1}%
  \BibitemOpen
  \bibfield  {author} {\bibinfo {author} {\bibfnamefont {Z.}~\bibnamefont
  {Romanowski}},\ }\href@noop {} {\bibfield  {journal} {\bibinfo  {journal}
  {Modell. Simul. Mater. Sci. Eng.}\ }\textbf {\bibinfo {volume} {16}},\
  \bibinfo {pages} {015003} (\bibinfo {year} {2007})}\BibitemShut {NoStop}%
\bibitem [{\citenamefont {Romanowski}(2009)}]{Romanowski2}%
  \BibitemOpen
  \bibfield  {author} {\bibinfo {author} {\bibfnamefont {Z.}~\bibnamefont
  {Romanowski}},\ }\href@noop {} {\bibfield  {journal} {\bibinfo  {journal}
  {Modell. Simul. Mater. Sci. Eng.}\ }\textbf {\bibinfo {volume} {17}},\
  \bibinfo {pages} {045001} (\bibinfo {year} {2009})}\BibitemShut {NoStop}%
\bibitem [{\citenamefont {Ozaki}\ and\ \citenamefont
  {Toyoda}(2011)}]{OZAKI20111245}%
  \BibitemOpen
  \bibfield  {author} {\bibinfo {author} {\bibfnamefont {T.}~\bibnamefont
  {Ozaki}}\ and\ \bibinfo {author} {\bibfnamefont {M.}~\bibnamefont {Toyoda}},\
  }\href@noop {} {\bibfield  {journal} {\bibinfo  {journal} {Comput. Phys.
  Commun.}\ }\textbf {\bibinfo {volume} {182}},\ \bibinfo {pages} {1245}
  (\bibinfo {year} {2011})}\BibitemShut {NoStop}%
\bibitem [{\citenamefont {Andrae}, \citenamefont {Brodbeck},\ and\
  \citenamefont {Hinze}(2001)}]{numerical_grid_atom}%
  \BibitemOpen
  \bibfield  {author} {\bibinfo {author} {\bibfnamefont {D.}~\bibnamefont
  {Andrae}}, \bibinfo {author} {\bibfnamefont {R.}~\bibnamefont {Brodbeck}}, \
  and\ \bibinfo {author} {\bibfnamefont {J.}~\bibnamefont {Hinze}},\
  }\href@noop {} {\bibfield  {journal} {\bibinfo  {journal} {Int. J. Quantum
  Chem.}\ }\textbf {\bibinfo {volume} {82}},\ \bibinfo {pages} {227} (\bibinfo
  {year} {2001})}\BibitemShut {NoStop}%
\bibitem [{\citenamefont {Herman}\ and\ \citenamefont
  {Skillman}(1963)}]{herman1963atomic}%
  \BibitemOpen
  \bibfield  {author} {\bibinfo {author} {\bibfnamefont {F.}~\bibnamefont
  {Herman}}\ and\ \bibinfo {author} {\bibfnamefont {S.}~\bibnamefont
  {Skillman}},\ }\href@noop {} {\emph {\bibinfo {title} {Atomic Structure
  Calculations}}}\ (\bibinfo  {publisher} {Prentice-Hall},\ \bibinfo {address}
  {Englewood Cliffs, NJ},\ \bibinfo {year} {1963})\BibitemShut {NoStop}%
\bibitem [{\citenamefont {Kotochigova}\ \emph {et~al.}(1997)\citenamefont
  {Kotochigova}, \citenamefont {Levine}, \citenamefont {Shirley}, \citenamefont
  {Stiles},\ and\ \citenamefont {Clark}}]{LDAexpatom}%
  \BibitemOpen
  \bibfield  {author} {\bibinfo {author} {\bibfnamefont {S.}~\bibnamefont
  {Kotochigova}}, \bibinfo {author} {\bibfnamefont {Z.~H.}\ \bibnamefont
  {Levine}}, \bibinfo {author} {\bibfnamefont {E.~L.}\ \bibnamefont {Shirley}},
  \bibinfo {author} {\bibfnamefont {M.~D.}\ \bibnamefont {Stiles}}, \ and\
  \bibinfo {author} {\bibfnamefont {C.~W.}\ \bibnamefont {Clark}},\ }\href@noop
  {} {\bibfield  {journal} {\bibinfo  {journal} {Phys. Rev. A}\ }\textbf
  {\bibinfo {volume} {55}},\ \bibinfo {pages} {191} (\bibinfo {year}
  {1997})}\BibitemShut {NoStop}%
\bibitem [{\citenamefont {Jiang}\ and\ \citenamefont
  {Engel}(2005)}]{secondorderKSMP2}%
  \BibitemOpen
  \bibfield  {author} {\bibinfo {author} {\bibfnamefont {H.}~\bibnamefont
  {Jiang}}\ and\ \bibinfo {author} {\bibfnamefont {E.}~\bibnamefont {Engel}},\
  }\href@noop {} {\bibfield  {journal} {\bibinfo  {journal} {J. Chem. Phys.}\
  }\textbf {\bibinfo {volume} {123}},\ \bibinfo {pages} {224102} (\bibinfo
  {year} {2005})}\BibitemShut {NoStop}%
\bibitem [{\citenamefont {Hellgren}\ and\ \citenamefont {von
  Barth}(2007)}]{gwasphericalatomshellgren}%
  \BibitemOpen
  \bibfield  {author} {\bibinfo {author} {\bibfnamefont {M.}~\bibnamefont
  {Hellgren}}\ and\ \bibinfo {author} {\bibfnamefont {U.}~\bibnamefont {von
  Barth}},\ }\href@noop {} {\bibfield  {journal} {\bibinfo  {journal} {Phys.
  Rev. B}\ }\textbf {\bibinfo {volume} {76}},\ \bibinfo {pages} {075107}
  (\bibinfo {year} {2007})}\BibitemShut {NoStop}%
\bibitem [{\citenamefont {Vacondio}\ \emph {et~al.}(2022)\citenamefont
  {Vacondio}, \citenamefont {Varsano}, \citenamefont {Ruini},\ and\
  \citenamefont {Ferretti}}]{vacondiopaper}%
  \BibitemOpen
  \bibfield  {author} {\bibinfo {author} {\bibfnamefont {S.}~\bibnamefont
  {Vacondio}}, \bibinfo {author} {\bibfnamefont {D.}~\bibnamefont {Varsano}},
  \bibinfo {author} {\bibfnamefont {A.}~\bibnamefont {Ruini}}, \ and\ \bibinfo
  {author} {\bibfnamefont {A.}~\bibnamefont {Ferretti}},\ }\href@noop {}
  {\bibfield  {journal} {\bibinfo  {journal} {J. Chem. Theory Comput.}\
  }\textbf {\bibinfo {volume} {18}},\ \bibinfo {pages} {3703} (\bibinfo {year}
  {2022})}\BibitemShut {NoStop}%
\bibitem [{\citenamefont {Užulis}\ and\ \citenamefont
  {Gulans}(2022)}]{Uzulis_2022}%
  \BibitemOpen
  \bibfield  {author} {\bibinfo {author} {\bibfnamefont {J.}~\bibnamefont
  {Užulis}}\ and\ \bibinfo {author} {\bibfnamefont {A.}~\bibnamefont
  {Gulans}},\ }\href@noop {} {\bibfield  {journal} {\bibinfo  {journal} {J.
  Phys. Commun.}\ }\textbf {\bibinfo {volume} {6}},\ \bibinfo {pages} {085002}
  (\bibinfo {year} {2022})}\BibitemShut {NoStop}%
\bibitem [{\citenamefont {Erhard}, \citenamefont {Trushin},\ and\ \citenamefont
  {Görling}(2022)}]{CCSD(T)inversion}%
  \BibitemOpen
  \bibfield  {author} {\bibinfo {author} {\bibfnamefont {J.}~\bibnamefont
  {Erhard}}, \bibinfo {author} {\bibfnamefont {E.}~\bibnamefont {Trushin}}, \
  and\ \bibinfo {author} {\bibfnamefont {A.}~\bibnamefont {Görling}},\
  }\href@noop {} {\bibfield  {journal} {\bibinfo  {journal} {J. Chem. Phys.}\
  }\textbf {\bibinfo {volume} {156}},\ \bibinfo {pages} {204124} (\bibinfo
  {year} {2022})}\BibitemShut {NoStop}%
\bibitem [{\citenamefont {Cao}, \citenamefont {Wang},\ and\ \citenamefont
  {Yang}(2017)}]{CCSD(T)openshellatoms}%
  \BibitemOpen
  \bibfield  {author} {\bibinfo {author} {\bibfnamefont {Z.}~\bibnamefont
  {Cao}}, \bibinfo {author} {\bibfnamefont {F.}~\bibnamefont {Wang}}, \ and\
  \bibinfo {author} {\bibfnamefont {M.}~\bibnamefont {Yang}},\ }\href@noop {}
  {\bibfield  {journal} {\bibinfo  {journal} {J. Chem. Phys.}\ }\textbf
  {\bibinfo {volume} {146}},\ \bibinfo {pages} {134108} (\bibinfo {year}
  {2017})}\BibitemShut {NoStop}%
\bibitem [{\citenamefont {Davidson}\ \emph {et~al.}(1991)\citenamefont
  {Davidson}, \citenamefont {Hagstrom}, \citenamefont {Chakravorty},
  \citenamefont {Umar},\ and\ \citenamefont {Fischer}}]{chakravarthyhelium}%
  \BibitemOpen
  \bibfield  {author} {\bibinfo {author} {\bibfnamefont {E.~R.}\ \bibnamefont
  {Davidson}}, \bibinfo {author} {\bibfnamefont {S.~A.}\ \bibnamefont
  {Hagstrom}}, \bibinfo {author} {\bibfnamefont {S.~J.}\ \bibnamefont
  {Chakravorty}}, \bibinfo {author} {\bibfnamefont {V.~M.}\ \bibnamefont
  {Umar}}, \ and\ \bibinfo {author} {\bibfnamefont {C.~F.}\ \bibnamefont
  {Fischer}},\ }\href@noop {} {\bibfield  {journal} {\bibinfo  {journal} {Phys.
  Rev. A}\ }\textbf {\bibinfo {volume} {44}},\ \bibinfo {pages} {7071}
  (\bibinfo {year} {1991})}\BibitemShut {NoStop}%
\bibitem [{\citenamefont {Chakravorty}\ \emph {et~al.}(1993)\citenamefont
  {Chakravorty}, \citenamefont {Gwaltney}, \citenamefont {Davidson},
  \citenamefont {Parpia},\ and\ \citenamefont
  {Fischer}}]{chakravarthyberylliumandothers}%
  \BibitemOpen
  \bibfield  {author} {\bibinfo {author} {\bibfnamefont {S.~J.}\ \bibnamefont
  {Chakravorty}}, \bibinfo {author} {\bibfnamefont {S.~R.}\ \bibnamefont
  {Gwaltney}}, \bibinfo {author} {\bibfnamefont {E.~R.}\ \bibnamefont
  {Davidson}}, \bibinfo {author} {\bibfnamefont {F.~A.}\ \bibnamefont
  {Parpia}}, \ and\ \bibinfo {author} {\bibfnamefont {C.~F.}\ \bibnamefont
  {Fischer}},\ }\href@noop {} {\bibfield  {journal} {\bibinfo  {journal} {Phys.
  Rev. A}\ }\textbf {\bibinfo {volume} {47}},\ \bibinfo {pages} {3649}
  (\bibinfo {year} {1993})}\BibitemShut {NoStop}%
\bibitem [{\citenamefont {Umrigar}\ and\ \citenamefont
  {Gonze}(1994)}]{umrigargonze}%
  \BibitemOpen
  \bibfield  {author} {\bibinfo {author} {\bibfnamefont {C.~J.}\ \bibnamefont
  {Umrigar}}\ and\ \bibinfo {author} {\bibfnamefont {X.}~\bibnamefont
  {Gonze}},\ }\href@noop {} {\bibfield  {journal} {\bibinfo  {journal} {Phys.
  Rev. A}\ }\textbf {\bibinfo {volume} {50}},\ \bibinfo {pages} {3827}
  (\bibinfo {year} {1994})}\BibitemShut {NoStop}%
\bibitem [{\citenamefont {Grabowski}\ \emph {et~al.}(2002)\citenamefont
  {Grabowski}, \citenamefont {Hirata}, \citenamefont {Ivanov},\ and\
  \citenamefont {Bartlett}}]{OEP-correlation-2-ivanov}%
  \BibitemOpen
  \bibfield  {author} {\bibinfo {author} {\bibfnamefont {I.}~\bibnamefont
  {Grabowski}}, \bibinfo {author} {\bibfnamefont {S.}~\bibnamefont {Hirata}},
  \bibinfo {author} {\bibfnamefont {S.}~\bibnamefont {Ivanov}}, \ and\ \bibinfo
  {author} {\bibfnamefont {R.~J.}\ \bibnamefont {Bartlett}},\ }\href@noop {}
  {\bibfield  {journal} {\bibinfo  {journal} {J. Chem. Phys.}\ }\textbf
  {\bibinfo {volume} {116}},\ \bibinfo {pages} {4415} (\bibinfo {year}
  {2002})}\BibitemShut {NoStop}%
\bibitem [{\citenamefont {Bartlett}\ \emph {et~al.}(2004)\citenamefont
  {Bartlett}, \citenamefont {Grabowski}, \citenamefont {Hirata},\ and\
  \citenamefont {Ivanov}}]{OEP-correlation-4-ivanov-sc2}%
  \BibitemOpen
  \bibfield  {author} {\bibinfo {author} {\bibfnamefont {R.~J.}\ \bibnamefont
  {Bartlett}}, \bibinfo {author} {\bibfnamefont {I.}~\bibnamefont {Grabowski}},
  \bibinfo {author} {\bibfnamefont {S.}~\bibnamefont {Hirata}}, \ and\ \bibinfo
  {author} {\bibfnamefont {S.}~\bibnamefont {Ivanov}},\ }\href@noop {}
  {\bibfield  {journal} {\bibinfo  {journal} {J. Chem. Phys.}\ }\textbf
  {\bibinfo {volume} {122}},\ \bibinfo {pages} {034104} (\bibinfo {year}
  {2004})}\BibitemShut {NoStop}%
\bibitem [{\citenamefont {Grabowski}\ \emph {et~al.}(2011)\citenamefont
  {Grabowski}, \citenamefont {Teale}, \citenamefont {Śmiga},\ and\
  \citenamefont {Bartlett}}]{OEP-correlation-5-grabowski}%
  \BibitemOpen
  \bibfield  {author} {\bibinfo {author} {\bibfnamefont {I.}~\bibnamefont
  {Grabowski}}, \bibinfo {author} {\bibfnamefont {A.~M.}\ \bibnamefont
  {Teale}}, \bibinfo {author} {\bibfnamefont {S.}~\bibnamefont {Śmiga}}, \
  and\ \bibinfo {author} {\bibfnamefont {R.~J.}\ \bibnamefont {Bartlett}},\
  }\href@noop {} {\bibfield  {journal} {\bibinfo  {journal} {J. Chem. Phys.}\
  }\textbf {\bibinfo {volume} {135}},\ \bibinfo {pages} {114111} (\bibinfo
  {year} {2011})}\BibitemShut {NoStop}%
\bibitem [{\citenamefont {Grabowski}(2008)}]{OEP-correlation-6-ivanov}%
  \BibitemOpen
  \bibfield  {author} {\bibinfo {author} {\bibfnamefont {I.}~\bibnamefont
  {Grabowski}},\ }\href@noop {} {\bibfield  {journal} {\bibinfo  {journal}
  {Int. J. Quantum Chem.}\ }\textbf {\bibinfo {volume} {108}},\ \bibinfo
  {pages} {2076} (\bibinfo {year} {2008})}\BibitemShut {NoStop}%
\bibitem [{\citenamefont {Śmiga}\ \emph {et~al.}(2020)\citenamefont {Śmiga},
  \citenamefont {Marusiak}, \citenamefont {Grabowski},\ and\ \citenamefont
  {Fabiano}}]{OEP-correlation-7-ivanov}%
  \BibitemOpen
  \bibfield  {author} {\bibinfo {author} {\bibfnamefont {S.}~\bibnamefont
  {Śmiga}}, \bibinfo {author} {\bibfnamefont {V.}~\bibnamefont {Marusiak}},
  \bibinfo {author} {\bibfnamefont {I.}~\bibnamefont {Grabowski}}, \ and\
  \bibinfo {author} {\bibfnamefont {E.}~\bibnamefont {Fabiano}},\ }\href@noop
  {} {\bibfield  {journal} {\bibinfo  {journal} {J. Chem. Phys.}\ }\textbf
  {\bibinfo {volume} {152}},\ \bibinfo {pages} {054109} (\bibinfo {year}
  {2020})}\BibitemShut {NoStop}%
\bibitem [{\citenamefont {Schweigert}, \citenamefont {Lotrich},\ and\
  \citenamefont {Bartlett}(2006)}]{OEP-correlation-8-Schweigert}%
  \BibitemOpen
  \bibfield  {author} {\bibinfo {author} {\bibfnamefont {I.~V.}\ \bibnamefont
  {Schweigert}}, \bibinfo {author} {\bibfnamefont {V.~F.}\ \bibnamefont
  {Lotrich}}, \ and\ \bibinfo {author} {\bibfnamefont {R.~J.}\ \bibnamefont
  {Bartlett}},\ }\href@noop {} {\bibfield  {journal} {\bibinfo  {journal} {J.
  Chem. Phys.}\ }\textbf {\bibinfo {volume} {125}},\ \bibinfo {pages} {104108}
  (\bibinfo {year} {2006})}\BibitemShut {NoStop}%
\bibitem [{\citenamefont {Verma}\ and\ \citenamefont
  {Bartlett}(2012)}]{vermabarlettpaper}%
  \BibitemOpen
  \bibfield  {author} {\bibinfo {author} {\bibfnamefont {P.}~\bibnamefont
  {Verma}}\ and\ \bibinfo {author} {\bibfnamefont {R.~J.}\ \bibnamefont
  {Bartlett}},\ }\href@noop {} {\bibfield  {journal} {\bibinfo  {journal} {J.
  Chem. Phys.}\ }\textbf {\bibinfo {volume} {136}},\ \bibinfo {pages} {044105}
  (\bibinfo {year} {2012})}\BibitemShut {NoStop}%
\bibitem [{\citenamefont {Hellgren}, \citenamefont {Rohr},\ and\ \citenamefont
  {Gross}(2012)}]{OEP-correlation-RPA-1-hellgren-molecules}%
  \BibitemOpen
  \bibfield  {author} {\bibinfo {author} {\bibfnamefont {M.}~\bibnamefont
  {Hellgren}}, \bibinfo {author} {\bibfnamefont {D.~R.}\ \bibnamefont {Rohr}},
  \ and\ \bibinfo {author} {\bibfnamefont {E.~K.~U.}\ \bibnamefont {Gross}},\
  }\href@noop {} {\bibfield  {journal} {\bibinfo  {journal} {J. Chem. Phys.}\
  }\textbf {\bibinfo {volume} {136}},\ \bibinfo {pages} {034106} (\bibinfo
  {year} {2012})}\BibitemShut {NoStop}%
\bibitem [{\citenamefont {Trushin}\ \emph {et~al.}(2025)\citenamefont
  {Trushin}, \citenamefont {Fauser}, \citenamefont {M\"olkner}, \citenamefont
  {Erhard},\ and\ \citenamefont {G\"orling}}]{scRPAgorlingpaper}%
  \BibitemOpen
  \bibfield  {author} {\bibinfo {author} {\bibfnamefont {E.}~\bibnamefont
  {Trushin}}, \bibinfo {author} {\bibfnamefont {S.}~\bibnamefont {Fauser}},
  \bibinfo {author} {\bibfnamefont {A.}~\bibnamefont {M\"olkner}}, \bibinfo
  {author} {\bibfnamefont {J.}~\bibnamefont {Erhard}}, \ and\ \bibinfo {author}
  {\bibfnamefont {A.}~\bibnamefont {G\"orling}},\ }\href@noop {} {\bibfield
  {journal} {\bibinfo  {journal} {Phys. Rev. Lett.}\ }\textbf {\bibinfo
  {volume} {134}},\ \bibinfo {pages} {016402} (\bibinfo {year}
  {2025})}\BibitemShut {NoStop}%
\bibitem [{\citenamefont {Polak}, \citenamefont {Zhao},\ and\ \citenamefont
  {Vuckovic}(2025)}]{realspacesecondorderML}%
  \BibitemOpen
  \bibfield  {author} {\bibinfo {author} {\bibfnamefont {E.}~\bibnamefont
  {Polak}}, \bibinfo {author} {\bibfnamefont {H.}~\bibnamefont {Zhao}}, \ and\
  \bibinfo {author} {\bibfnamefont {S.}~\bibnamefont {Vuckovic}},\ }\href@noop
  {} {\bibfield  {journal} {\bibinfo  {journal} {Nat. Commun.}\ } (\bibinfo
  {year} {2025})}\BibitemShut {NoStop}%
\bibitem [{\citenamefont {Bilous}, \citenamefont {P\'alffy},\ and\
  \citenamefont {Marquardt}(2023)}]{MLCI}%
  \BibitemOpen
  \bibfield  {author} {\bibinfo {author} {\bibfnamefont {P.}~\bibnamefont
  {Bilous}}, \bibinfo {author} {\bibfnamefont {A.}~\bibnamefont {P\'alffy}}, \
  and\ \bibinfo {author} {\bibfnamefont {F.}~\bibnamefont {Marquardt}},\
  }\href@noop {} {\bibfield  {journal} {\bibinfo  {journal} {Phys. Rev. Lett.}\
  }\textbf {\bibinfo {volume} {131}},\ \bibinfo {pages} {133002} (\bibinfo
  {year} {2023})}\BibitemShut {NoStop}%
\bibitem [{\citenamefont {Cinal}(2020)}]{cinal}%
  \BibitemOpen
  \bibfield  {author} {\bibinfo {author} {\bibfnamefont {M.}~\bibnamefont
  {Cinal}},\ }\href@noop {} {\bibfield  {journal} {\bibinfo  {journal} {J.
  Math. Chem.}\ }\textbf {\bibinfo {volume} {58}} (\bibinfo {year}
  {2020})}\BibitemShut {NoStop}%
\bibitem [{\citenamefont {Jiang}\ and\ \citenamefont
  {Engel}(2007)}]{RPAcorrelationenergyjiangengel}%
  \BibitemOpen
  \bibfield  {author} {\bibinfo {author} {\bibfnamefont {H.}~\bibnamefont
  {Jiang}}\ and\ \bibinfo {author} {\bibfnamefont {E.}~\bibnamefont {Engel}},\
  }\href@noop {} {\bibfield  {journal} {\bibinfo  {journal} {J. Chem. Phys.}\
  }\textbf {\bibinfo {volume} {127}},\ \bibinfo {pages} {184108} (\bibinfo
  {year} {2007})}\BibitemShut {NoStop}%
\bibitem [{\citenamefont {K\"ummel}\ and\ \citenamefont
  {Perdew}(2003)}]{OEP-exchange-4-kummel-atoms-molecules}%
  \BibitemOpen
  \bibfield  {author} {\bibinfo {author} {\bibfnamefont {S.}~\bibnamefont
  {K\"ummel}}\ and\ \bibinfo {author} {\bibfnamefont {J.~P.}\ \bibnamefont
  {Perdew}},\ }\href@noop {} {\bibfield  {journal} {\bibinfo  {journal} {Phys.
  Rev. Lett.}\ }\textbf {\bibinfo {volume} {90}},\ \bibinfo {pages} {043004}
  (\bibinfo {year} {2003})}\BibitemShut {NoStop}%
\bibitem [{\citenamefont {G\"orling}\ and\ \citenamefont
  {Levy}(1994)}]{OEP-correlation-GL2-2}%
  \BibitemOpen
  \bibfield  {author} {\bibinfo {author} {\bibfnamefont {A.}~\bibnamefont
  {G\"orling}}\ and\ \bibinfo {author} {\bibfnamefont {M.}~\bibnamefont
  {Levy}},\ }\href@noop {} {\bibfield  {journal} {\bibinfo  {journal} {Phys.
  Rev. A}\ }\textbf {\bibinfo {volume} {50}},\ \bibinfo {pages} {196} (\bibinfo
  {year} {1994})}\BibitemShut {NoStop}%
\bibitem [{\citenamefont {Engel}\ and\ \citenamefont
  {Dreizler}(2011)}]{engel_dreizler_2011}%
  \BibitemOpen
  \bibfield  {author} {\bibinfo {author} {\bibfnamefont {E.}~\bibnamefont
  {Engel}}\ and\ \bibinfo {author} {\bibfnamefont {R.~M.}\ \bibnamefont
  {Dreizler}},\ }\href@noop {} {\emph {\bibinfo {title} {Density functional
  theory: An advanced course}}}\ (\bibinfo  {publisher} {Springer},\ \bibinfo
  {year} {2011})\BibitemShut {NoStop}%
\bibitem [{\citenamefont {Patera}(1984)}]{PATERA1984468}%
  \BibitemOpen
  \bibfield  {author} {\bibinfo {author} {\bibfnamefont {A.~T.}\ \bibnamefont
  {Patera}},\ }\href@noop {} {\bibfield  {journal} {\bibinfo  {journal} {J.
  Comput. Phys.}\ }\textbf {\bibinfo {volume} {54}},\ \bibinfo {pages} {468}
  (\bibinfo {year} {1984})}\BibitemShut {NoStop}%
\bibitem [{\citenamefont {Aashamar}, \citenamefont {Luke},\ and\ \citenamefont
  {Talman}(1979)}]{OEP-talman-5}%
  \BibitemOpen
  \bibfield  {author} {\bibinfo {author} {\bibfnamefont {K.}~\bibnamefont
  {Aashamar}}, \bibinfo {author} {\bibfnamefont {T.~M.}\ \bibnamefont {Luke}},
  \ and\ \bibinfo {author} {\bibfnamefont {J.~D.}\ \bibnamefont {Talman}},\
  }\href@noop {} {\bibfield  {journal} {\bibinfo  {journal} {Phys. Rev. A}\
  }\textbf {\bibinfo {volume} {19}},\ \bibinfo {pages} {6} (\bibinfo {year}
  {1979})}\BibitemShut {NoStop}%
\bibitem [{\citenamefont {Norman}\ and\ \citenamefont
  {Koelling}(1984)}]{GoodOEPaper}%
  \BibitemOpen
  \bibfield  {author} {\bibinfo {author} {\bibfnamefont {M.~R.}\ \bibnamefont
  {Norman}}\ and\ \bibinfo {author} {\bibfnamefont {D.~D.}\ \bibnamefont
  {Koelling}},\ }\href@noop {} {\bibfield  {journal} {\bibinfo  {journal}
  {Phys. Rev. B}\ }\textbf {\bibinfo {volume} {30}},\ \bibinfo {pages} {5530}
  (\bibinfo {year} {1984})}\BibitemShut {NoStop}%
\bibitem [{\citenamefont {Pask}\ and\ \citenamefont
  {Sterne}(2005)}]{pask2005finite}%
  \BibitemOpen
  \bibfield  {author} {\bibinfo {author} {\bibfnamefont {J.}~\bibnamefont
  {Pask}}\ and\ \bibinfo {author} {\bibfnamefont {P.}~\bibnamefont {Sterne}},\
  }\href@noop {} {\bibfield  {journal} {\bibinfo  {journal} {Modell. Simul.
  Mater. Sci. Eng.}\ }\textbf {\bibinfo {volume} {13}},\ \bibinfo {pages} {R71}
  (\bibinfo {year} {2005})}\BibitemShut {NoStop}%
\bibitem [{\citenamefont {Suryanarayana}\ \emph {et~al.}(2010)\citenamefont
  {Suryanarayana}, \citenamefont {Gavini}, \citenamefont {Blesgen},
  \citenamefont {Bhattacharya},\ and\ \citenamefont
  {Ortiz}}]{SURYANARAYANA2010256}%
  \BibitemOpen
  \bibfield  {author} {\bibinfo {author} {\bibfnamefont {P.}~\bibnamefont
  {Suryanarayana}}, \bibinfo {author} {\bibfnamefont {V.}~\bibnamefont
  {Gavini}}, \bibinfo {author} {\bibfnamefont {T.}~\bibnamefont {Blesgen}},
  \bibinfo {author} {\bibfnamefont {K.}~\bibnamefont {Bhattacharya}}, \ and\
  \bibinfo {author} {\bibfnamefont {M.}~\bibnamefont {Ortiz}},\ }\href@noop {}
  {\bibfield  {journal} {\bibinfo  {journal} {J. Mech. Phys. Solids}\ }\textbf
  {\bibinfo {volume} {58}},\ \bibinfo {pages} {256} (\bibinfo {year}
  {2010})}\BibitemShut {NoStop}%
\bibitem [{\citenamefont {Motamarri}\ \emph {et~al.}(2020)\citenamefont
  {Motamarri}, \citenamefont {Das}, \citenamefont {Rudraraju}, \citenamefont
  {Ghosh}, \citenamefont {Davydov},\ and\ \citenamefont
  {Gavini}}]{DFTFEVIKRAM}%
  \BibitemOpen
  \bibfield  {author} {\bibinfo {author} {\bibfnamefont {P.}~\bibnamefont
  {Motamarri}}, \bibinfo {author} {\bibfnamefont {S.}~\bibnamefont {Das}},
  \bibinfo {author} {\bibfnamefont {S.}~\bibnamefont {Rudraraju}}, \bibinfo
  {author} {\bibfnamefont {K.}~\bibnamefont {Ghosh}}, \bibinfo {author}
  {\bibfnamefont {D.}~\bibnamefont {Davydov}}, \ and\ \bibinfo {author}
  {\bibfnamefont {V.}~\bibnamefont {Gavini}},\ }\href@noop {} {\bibfield
  {journal} {\bibinfo  {journal} {Comput. Phys. Commun.}\ }\textbf {\bibinfo
  {volume} {246}},\ \bibinfo {pages} {106853} (\bibinfo {year}
  {2020})}\BibitemShut {NoStop}%
\bibitem [{\citenamefont {Ivanov}, \citenamefont {Hirata},\ and\ \citenamefont
  {Bartlett}(2002)}]{OEP-singularity-2-ivanov}%
  \BibitemOpen
  \bibfield  {author} {\bibinfo {author} {\bibfnamefont {S.}~\bibnamefont
  {Ivanov}}, \bibinfo {author} {\bibfnamefont {S.}~\bibnamefont {Hirata}}, \
  and\ \bibinfo {author} {\bibfnamefont {R.~J.}\ \bibnamefont {Bartlett}},\
  }\href@noop {} {\bibfield  {journal} {\bibinfo  {journal} {J. Chem. Phys.}\
  }\textbf {\bibinfo {volume} {116}},\ \bibinfo {pages} {1269} (\bibinfo {year}
  {2002})}\BibitemShut {NoStop}%
\bibitem [{\citenamefont {Banerjee}, \citenamefont {Suryanarayana},\ and\
  \citenamefont {Pask}(2016)}]{BANERJEE201631}%
  \BibitemOpen
  \bibfield  {author} {\bibinfo {author} {\bibfnamefont {A.~S.}\ \bibnamefont
  {Banerjee}}, \bibinfo {author} {\bibfnamefont {P.}~\bibnamefont
  {Suryanarayana}}, \ and\ \bibinfo {author} {\bibfnamefont {J.~E.}\
  \bibnamefont {Pask}},\ }\href@noop {} {\bibfield  {journal} {\bibinfo
  {journal} {Chem. Phys. Lett.}\ }\textbf {\bibinfo {volume} {647}},\ \bibinfo
  {pages} {31} (\bibinfo {year} {2016})}\BibitemShut {NoStop}%
\bibitem [{\citenamefont {Peng}\ \emph {et~al.}(2023)\citenamefont {Peng},
  \citenamefont {Yang}, \citenamefont {Jiang}, \citenamefont {Weng},\ and\
  \citenamefont {Ren}}]{nscRPA_sternheimer_atoms}%
  \BibitemOpen
  \bibfield  {author} {\bibinfo {author} {\bibfnamefont {H.}~\bibnamefont
  {Peng}}, \bibinfo {author} {\bibfnamefont {S.}~\bibnamefont {Yang}}, \bibinfo
  {author} {\bibfnamefont {H.}~\bibnamefont {Jiang}}, \bibinfo {author}
  {\bibfnamefont {H.}~\bibnamefont {Weng}}, \ and\ \bibinfo {author}
  {\bibfnamefont {X.}~\bibnamefont {Ren}},\ }\href@noop {} {\bibfield
  {journal} {\bibinfo  {journal} {J. Chem. Theory Comput.}\ }\textbf {\bibinfo
  {volume} {19}},\ \bibinfo {pages} {7199} (\bibinfo {year}
  {2023})}\BibitemShut {NoStop}%
\bibitem [{\citenamefont {Peng}\ and\ \citenamefont
  {Ren}(2025)}]{nscRPA_sternheimer_molecules}%
  \BibitemOpen
  \bibfield  {author} {\bibinfo {author} {\bibfnamefont {H.}~\bibnamefont
  {Peng}}\ and\ \bibinfo {author} {\bibfnamefont {X.}~\bibnamefont {Ren}},\
  }\href@noop {} {\bibfield  {journal} {\bibinfo  {journal} {Phys. Rev. A}\
  }\textbf {\bibinfo {volume} {112}},\ \bibinfo {pages} {062819} (\bibinfo
  {year} {2025})}\BibitemShut {NoStop}%
\bibitem [{\citenamefont {Peng}\ \emph {et~al.}(2025)\citenamefont {Peng},
  \citenamefont {Liu}, \citenamefont {Li}, \citenamefont {Xie},\ and\
  \citenamefont {Ren}}]{nscRPA_delta_FE_molecules}%
  \BibitemOpen
  \bibfield  {author} {\bibinfo {author} {\bibfnamefont {H.}~\bibnamefont
  {Peng}}, \bibinfo {author} {\bibfnamefont {H.}~\bibnamefont {Liu}}, \bibinfo
  {author} {\bibfnamefont {C.}~\bibnamefont {Li}}, \bibinfo {author}
  {\bibfnamefont {H.}~\bibnamefont {Xie}}, \ and\ \bibinfo {author}
  {\bibfnamefont {X.}~\bibnamefont {Ren}},\ }\href@noop {} {\enquote {\bibinfo
  {title} {A finite-element delta-sternheimer approach for accurate
  all-electron rpa correlation energies of arbitrary molecules},}\ } (\bibinfo
  {year} {2025}),\ \Eprint {http://arxiv.org/abs/2510.15570} {arXiv:2510.15570}
  \BibitemShut {NoStop}%
\bibitem [{\citenamefont {Kurth}\ and\ \citenamefont
  {Perdew}(1999)}]{Perdewkurthrpaplus}%
  \BibitemOpen
  \bibfield  {author} {\bibinfo {author} {\bibfnamefont {S.}~\bibnamefont
  {Kurth}}\ and\ \bibinfo {author} {\bibfnamefont {J.~P.}\ \bibnamefont
  {Perdew}},\ }\href@noop {} {\bibfield  {journal} {\bibinfo  {journal} {Phys.
  Rev. B}\ }\textbf {\bibinfo {volume} {59}},\ \bibinfo {pages} {10461}
  (\bibinfo {year} {1999})}\BibitemShut {NoStop}%
\bibitem [{\citenamefont {Yan}, \citenamefont {Perdew},\ and\ \citenamefont
  {Kurth}(2000)}]{perdewzidankurth}%
  \BibitemOpen
  \bibfield  {author} {\bibinfo {author} {\bibfnamefont {Z.}~\bibnamefont
  {Yan}}, \bibinfo {author} {\bibfnamefont {J.~P.}\ \bibnamefont {Perdew}}, \
  and\ \bibinfo {author} {\bibfnamefont {S.}~\bibnamefont {Kurth}},\
  }\href@noop {} {\bibfield  {journal} {\bibinfo  {journal} {Phys. Rev. B}\
  }\textbf {\bibinfo {volume} {61}},\ \bibinfo {pages} {16430} (\bibinfo {year}
  {2000})}\BibitemShut {NoStop}%
\bibitem [{\citenamefont {Ruzsinszky}, \citenamefont {Perdew},\ and\
  \citenamefont {Csonka}(2010)}]{RPA++perdew}%
  \BibitemOpen
  \bibfield  {author} {\bibinfo {author} {\bibfnamefont {A.}~\bibnamefont
  {Ruzsinszky}}, \bibinfo {author} {\bibfnamefont {J.~P.}\ \bibnamefont
  {Perdew}}, \ and\ \bibinfo {author} {\bibfnamefont {G.~I.}\ \bibnamefont
  {Csonka}},\ }\href@noop {} {\bibfield  {journal} {\bibinfo  {journal} {J.
  Chem. Theory Comput.}\ }\textbf {\bibinfo {volume} {6}},\ \bibinfo {pages}
  {127} (\bibinfo {year} {2010})}\BibitemShut {NoStop}%
\bibitem [{\citenamefont {Grimme}\ and\ \citenamefont
  {Steinmetz}(2016)}]{double-hybrid-RPA-1}%
  \BibitemOpen
  \bibfield  {author} {\bibinfo {author} {\bibfnamefont {S.}~\bibnamefont
  {Grimme}}\ and\ \bibinfo {author} {\bibfnamefont {M.}~\bibnamefont
  {Steinmetz}},\ }\href@noop {} {\bibfield  {journal} {\bibinfo  {journal}
  {Phys. Chem. Chem. Phys.}\ }\textbf {\bibinfo {volume} {18}},\ \bibinfo
  {pages} {20926} (\bibinfo {year} {2016})}\BibitemShut {NoStop}%
\bibitem [{\citenamefont {Mezei}\ \emph {et~al.}(2015)\citenamefont {Mezei},
  \citenamefont {Csonka}, \citenamefont {Ruzsinszky},\ and\ \citenamefont
  {Kállay}}]{double-hybrid-RPA-3}%
  \BibitemOpen
  \bibfield  {author} {\bibinfo {author} {\bibfnamefont {P.~D.}\ \bibnamefont
  {Mezei}}, \bibinfo {author} {\bibfnamefont {G.~I.}\ \bibnamefont {Csonka}},
  \bibinfo {author} {\bibfnamefont {A.}~\bibnamefont {Ruzsinszky}}, \ and\
  \bibinfo {author} {\bibfnamefont {M.}~\bibnamefont {Kállay}},\ }\href@noop
  {} {\bibfield  {journal} {\bibinfo  {journal} {J. Chem. Theory Comput.}\
  }\textbf {\bibinfo {volume} {11}},\ \bibinfo {pages} {4615} (\bibinfo {year}
  {2015})}\BibitemShut {NoStop}%
\bibitem [{\citenamefont {Brémond}\ and\ \citenamefont
  {Adamo}(2011)}]{double-hybrid-equation-1}%
  \BibitemOpen
  \bibfield  {author} {\bibinfo {author} {\bibfnamefont {E.}~\bibnamefont
  {Brémond}}\ and\ \bibinfo {author} {\bibfnamefont {C.}~\bibnamefont
  {Adamo}},\ }\href@noop {} {\bibfield  {journal} {\bibinfo  {journal} {J.
  Chem. Phys.}\ }\textbf {\bibinfo {volume} {135}},\ \bibinfo {pages} {024106}
  (\bibinfo {year} {2011})}\BibitemShut {NoStop}%
\bibitem [{\citenamefont {Perdew}, \citenamefont {Burke},\ and\ \citenamefont
  {Ernzerhof}(1996)}]{PBE}%
  \BibitemOpen
  \bibfield  {author} {\bibinfo {author} {\bibfnamefont {J.~P.}\ \bibnamefont
  {Perdew}}, \bibinfo {author} {\bibfnamefont {K.}~\bibnamefont {Burke}}, \
  and\ \bibinfo {author} {\bibfnamefont {M.}~\bibnamefont {Ernzerhof}},\
  }\href@noop {} {\bibfield  {journal} {\bibinfo  {journal} {Phys. Rev. Lett.}\
  }\textbf {\bibinfo {volume} {77}},\ \bibinfo {pages} {3865} (\bibinfo {year}
  {1996})}\BibitemShut {NoStop}%
\bibitem [{\citenamefont {Nagai}, \citenamefont {Akashi},\ and\ \citenamefont
  {Sugino}(2019)}]{Nagai2019CompletingDF}%
  \BibitemOpen
  \bibfield  {author} {\bibinfo {author} {\bibfnamefont {R.}~\bibnamefont
  {Nagai}}, \bibinfo {author} {\bibfnamefont {R.}~\bibnamefont {Akashi}}, \
  and\ \bibinfo {author} {\bibfnamefont {O.}~\bibnamefont {Sugino}},\
  }\href@noop {} {\bibfield  {journal} {\bibinfo  {journal} {npj Comput.
  Mater.}\ }\textbf {\bibinfo {volume} {6}},\ \bibinfo {pages} {1} (\bibinfo
  {year} {2019})}\BibitemShut {NoStop}%
\bibitem [{\citenamefont {Pokharel}\ \emph {et~al.}(2022)\citenamefont
  {Pokharel}, \citenamefont {Furness}, \citenamefont {Yao}, \citenamefont
  {Blum}, \citenamefont {Irons}, \citenamefont {Teale},\ and\ \citenamefont
  {Sun}}]{MLreducedgradientlaplacian}%
  \BibitemOpen
  \bibfield  {author} {\bibinfo {author} {\bibfnamefont {K.}~\bibnamefont
  {Pokharel}}, \bibinfo {author} {\bibfnamefont {J.~W.}\ \bibnamefont
  {Furness}}, \bibinfo {author} {\bibfnamefont {Y.}~\bibnamefont {Yao}},
  \bibinfo {author} {\bibfnamefont {V.}~\bibnamefont {Blum}}, \bibinfo {author}
  {\bibfnamefont {T.~J.~P.}\ \bibnamefont {Irons}}, \bibinfo {author}
  {\bibfnamefont {A.~M.}\ \bibnamefont {Teale}}, \ and\ \bibinfo {author}
  {\bibfnamefont {J.}~\bibnamefont {Sun}},\ }\href@noop {} {\bibfield
  {journal} {\bibinfo  {journal} {J. Chem. Phys.}\ }\textbf {\bibinfo {volume}
  {157}},\ \bibinfo {pages} {174106} (\bibinfo {year} {2022})}\BibitemShut
  {NoStop}%
\bibitem [{\citenamefont {Zheng}\ \emph {et~al.}(2025)\citenamefont {Zheng},
  \citenamefont {Zhou}, \citenamefont {Zhu}, \citenamefont {Zhuang},
  \citenamefont {Yam}, \citenamefont {Chen}, \citenamefont {An}, \citenamefont
  {Zheng}, \citenamefont {Hu},\ and\ \citenamefont {Chen}}]{Reducedgradient3}%
  \BibitemOpen
  \bibfield  {author} {\bibinfo {author} {\bibfnamefont {Y.}~\bibnamefont
  {Zheng}}, \bibinfo {author} {\bibfnamefont {Y.}~\bibnamefont {Zhou}},
  \bibinfo {author} {\bibfnamefont {Y.}~\bibnamefont {Zhu}}, \bibinfo {author}
  {\bibfnamefont {Y.}~\bibnamefont {Zhuang}}, \bibinfo {author} {\bibfnamefont
  {C.}~\bibnamefont {Yam}}, \bibinfo {author} {\bibfnamefont {Z.-H.}\
  \bibnamefont {Chen}}, \bibinfo {author} {\bibfnamefont {Z.}~\bibnamefont
  {An}}, \bibinfo {author} {\bibfnamefont {X.}~\bibnamefont {Zheng}}, \bibinfo
  {author} {\bibfnamefont {Z.}~\bibnamefont {Hu}}, \ and\ \bibinfo {author}
  {\bibfnamefont {G.}~\bibnamefont {Chen}},\ }\href@noop {} {\bibfield
  {journal} {\bibinfo  {journal} {J. Chem. Phys.}\ }\textbf {\bibinfo {volume}
  {163}},\ \bibinfo {pages} {184103} (\bibinfo {year} {2025})}\BibitemShut
  {NoStop}%
\bibitem [{\citenamefont {Kanungo}\ \emph {et~al.}(2025)\citenamefont
  {Kanungo}, \citenamefont {Hatch}, \citenamefont {Zimmerman},\ and\
  \citenamefont {Gavini}}]{MLvikram}%
  \BibitemOpen
  \bibfield  {author} {\bibinfo {author} {\bibfnamefont {B.}~\bibnamefont
  {Kanungo}}, \bibinfo {author} {\bibfnamefont {J.}~\bibnamefont {Hatch}},
  \bibinfo {author} {\bibfnamefont {P.~M.}\ \bibnamefont {Zimmerman}}, \ and\
  \bibinfo {author} {\bibfnamefont {V.}~\bibnamefont {Gavini}},\ }\href@noop {}
  {\bibfield  {journal} {\bibinfo  {journal} {Sci. Adv.}\ }\textbf {\bibinfo
  {volume} {11}},\ \bibinfo {pages} {eady8962} (\bibinfo {year}
  {2025})}\BibitemShut {NoStop}%
\bibitem [{\citenamefont {Kumar}\ \emph {et~al.}(2023)\citenamefont {Kumar},
  \citenamefont {Jing}, \citenamefont {Pask}, \citenamefont {Medford},\ and\
  \citenamefont {Suryanarayana}}]{deltalearning1}%
  \BibitemOpen
  \bibfield  {author} {\bibinfo {author} {\bibfnamefont {S.}~\bibnamefont
  {Kumar}}, \bibinfo {author} {\bibfnamefont {X.}~\bibnamefont {Jing}},
  \bibinfo {author} {\bibfnamefont {J.~E.}\ \bibnamefont {Pask}}, \bibinfo
  {author} {\bibfnamefont {A.~J.}\ \bibnamefont {Medford}}, \ and\ \bibinfo
  {author} {\bibfnamefont {P.}~\bibnamefont {Suryanarayana}},\ }\href@noop {}
  {\bibfield  {journal} {\bibinfo  {journal} {J. Chem. Phys.}\ }\textbf
  {\bibinfo {volume} {159}},\ \bibinfo {pages} {244106} (\bibinfo {year}
  {2023})}\BibitemShut {NoStop}%
\bibitem [{\citenamefont {Kumar}\ \emph {et~al.}(2024)\citenamefont {Kumar},
  \citenamefont {Jing}, \citenamefont {Pask},\ and\ \citenamefont
  {Suryanarayana}}]{MLFFwarmdensematter}%
  \BibitemOpen
  \bibfield  {author} {\bibinfo {author} {\bibfnamefont {S.}~\bibnamefont
  {Kumar}}, \bibinfo {author} {\bibfnamefont {X.}~\bibnamefont {Jing}},
  \bibinfo {author} {\bibfnamefont {J.~E.}\ \bibnamefont {Pask}}, \ and\
  \bibinfo {author} {\bibfnamefont {P.}~\bibnamefont {Suryanarayana}},\
  }\href@noop {} {\bibfield  {journal} {\bibinfo  {journal} {Phys. Plasmas}\
  }\textbf {\bibinfo {volume} {31}},\ \bibinfo {pages} {043905} (\bibinfo
  {year} {2024})}\BibitemShut {NoStop}%
\bibitem [{\citenamefont {Bishop}(2006)}]{bishopML}%
  \BibitemOpen
  \bibfield  {author} {\bibinfo {author} {\bibfnamefont {C.~M.}\ \bibnamefont
  {Bishop}},\ }\href@noop {} {\emph {\bibinfo {title} {Pattern Recognition and
  Machine Learning}}}\ (\bibinfo  {publisher} {Springer},\ \bibinfo {year}
  {2006})\BibitemShut {NoStop}%
\bibitem [{\citenamefont {Furness}\ \emph {et~al.}(2020)\citenamefont
  {Furness}, \citenamefont {Kaplan}, \citenamefont {Ning}, \citenamefont
  {Perdew},\ and\ \citenamefont {Sun}}]{r2SCAN}%
  \BibitemOpen
  \bibfield  {author} {\bibinfo {author} {\bibfnamefont {J.~W.}\ \bibnamefont
  {Furness}}, \bibinfo {author} {\bibfnamefont {A.~D.}\ \bibnamefont {Kaplan}},
  \bibinfo {author} {\bibfnamefont {J.}~\bibnamefont {Ning}}, \bibinfo {author}
  {\bibfnamefont {J.~P.}\ \bibnamefont {Perdew}}, \ and\ \bibinfo {author}
  {\bibfnamefont {J.}~\bibnamefont {Sun}},\ }\href@noop {} {\bibfield
  {journal} {\bibinfo  {journal} {J. Phys. Chem. Lett.}\ }\textbf {\bibinfo
  {volume} {11}},\ \bibinfo {pages} {8208} (\bibinfo {year}
  {2020})}\BibitemShut {NoStop}%
\bibitem [{\citenamefont {Riemelmoser}\ \emph {et~al.}(2023)\citenamefont
  {Riemelmoser}, \citenamefont {Verdi}, \citenamefont {Kaltak},\ and\
  \citenamefont {Kresse}}]{MLRPAKresse}%
  \BibitemOpen
  \bibfield  {author} {\bibinfo {author} {\bibfnamefont {S.}~\bibnamefont
  {Riemelmoser}}, \bibinfo {author} {\bibfnamefont {C.}~\bibnamefont {Verdi}},
  \bibinfo {author} {\bibfnamefont {M.}~\bibnamefont {Kaltak}}, \ and\ \bibinfo
  {author} {\bibfnamefont {G.}~\bibnamefont {Kresse}},\ }\href@noop {}
  {\bibfield  {journal} {\bibinfo  {journal} {J. Chem. Theory Comput.}\
  }\textbf {\bibinfo {volume} {19}},\ \bibinfo {pages} {7287} (\bibinfo {year}
  {2023})}\BibitemShut {NoStop}%
\end{thebibliography}
%


\end{document}